\documentclass[a4paper,fleqn,usenatbib]{mnras}

\usepackage{mathptmx}

\usepackage[T1]{fontenc}
\usepackage{ae,aecompl}

\pdfoutput=1

\usepackage{graphicx}	
\usepackage{amsmath}	
\usepackage{amssymb}	
\usepackage{hyperref}     
\usepackage[usenames, dvipsnames]{color} 

\newcommand{\pcmc}{cm$^{-3}$\,}	
\newcommand{\kms}{km\,s$^{-1}$\,}	

\graphicspath{{./plots/}}

\title[Resolution requirements to model molecular gas]{On the resolution requirements for modelling molecular gas formation in solar neighbourhood conditions}

\author[P. R. Joshi et al.]{
P. R. Joshi$^{1}$\thanks{E-mail: joshi@ph1.uni-koeln.de},
S. Walch$^{1}$\thanks{E-mail: walch@ph1.uni-koeln.de},
D. Seifried$^{1},$
S. C. O. Glover$^{2},$
S. D. Clarke$^{1},$
\& M. Weis$^{1}$
\\
$^{1}$Universit{\"a}t zu K{\"o}ln, I. Physikalisches Institut, Z{\"u}lpicher Str. 77, 50937 K{\"o}ln, Germany \\
$^{2}$Zentrum f{\"u}r  Astronomie der Universit{\"a}t Heidelberg, Institut f{\"u}r Theoretische Astrophysik, Albert-Ueberle-Str. 2, 69120 Heidelberg, Germany
}

\date{Accepted XXX. Received YYY; in original form ZZZ}

\pubyear{2018}

\begin{document}
\label{firstpage}
\pagerange{\pageref{firstpage}--\pageref{lastpage}}
\maketitle

\begin{abstract}
The formation of molecular hydrogen (H$_2$) and carbon monoxide (CO) is sensitive to the volume and column density distribution of the turbulent interstellar medium. In this paper, we study H$_2$ and CO formation in a large set of hydrodynamical simulations of periodic boxes with driven supersonic turbulence, as well as in colliding flows with the \textsc{Flash} code. The simulations include a non-equilibrium chemistry network, gas self-gravity, and diffuse radiative transfer. We investigate the spatial resolution required to obtain a converged H$_2$ and CO mass fraction and formation history. From the numerical tests we find that H$_2$ converges at a spatial resolution of $\lesssim0.2$~pc, while the required resolution for CO convergence is $\lesssim 0.04$~pc in gas with solar metallicity which is subject to a solar neighbourhood interstellar radiation field. We derive two critical conditions from our numerical results: the simulation has to at least resolve the densities at which (1) the molecule formation time in each cell in the computational domain is equal to the dissociation time, and (2) the formation time is equal to the the typical cell crossing time. For both H$_2$ and CO, the second criterion is more restrictive. The formulae we derive can be used to check whether molecule formation is converged in any given simulation.
\end{abstract}



\section{Introduction}
		\label{sec:intro}
		\vspace{3mm}
		Molecular clouds (MCs) are turbulent structures in the interstellar medium \citep[][]{Scalo04,MacLow04}, rich in chemical species that are not in chemical equilibrium \citep{Leung1984}. The overall density structure of the MC is known to significantly affect the chemical composition since it influences the shielding experienced by various regions within the cloud \citep[e.g.][]{Stutzki1990, Shrader17}. \citet{Glover10} suggest that the MC composition is dependent on the history of the gas since the dynamical time-scale and the molecule formation time-scale are comparable in MC conditions. Thus, simulations of MCs should couple the dynamical and chemical evolution to obtain correct approximations of the MC formation in nature. In doing so, the chemical abundances have to be evaluated at every hydrodynamical time-step; thus, only simple chemical networks are affordable. \citet{Koyama00, Koyama02} and \citet{Bergin04} implemented the non-equilibrium chemistry of H$_2$ and CO in their one- and two-dimensional hydrodynamic simulations because CO is the most important tracer of H$_2$ and an essential coolant in the local interstellar medium (ISM). With the increase in computational capabilities, chemistry networks with more species and reactions were used in three-dimensional magneto-hydrodynamic (MHD) simulations \citep[e.g.][]{Glover07a, Glover07b}. Later, \citet{Glover10} presented a chemistry network with 32 species linked by 218 reactions, which allows for an accurate treatment of H$_2$ and CO formation. A simpler, computationally cheaper network, based  on \citet[][]{Glover07a, Glover07b} and \citet{NL97} has been applied in many recent high-resolution (M)HD simulations \citep[e.g.][]{Walch11,Micic12,Walch15_SILCC1, Girichidis2016_SILCC2, Gatto2017_SILCC3, Pardi2017, Peters17_SILCC4, Seifried2017b}. Similarly, a few works \citep[e.g.][]{Glover2012chemcompare,Clark2012collflow,Szucs2014} have used a more complex network (termed ``NL99") that combines the CO chemistry of \citet{NL99} with the hydrogen chemistry of \citet[][]{Glover07a, Glover07b}. The NL99 network is used in the work presented here. Alternative, simplified networks for H$_2$ and/or CO formation have also been presented by \citet{Keto2008}, \citet{Valdivia2016}, and \citet{Gong2017}. In simulations of star-forming filaments, \citet{Seifried2016} presented the use of a heavier network containing 287 reactions between 37 chemical species, implemented as part of the \textsc{Krome} package \citep{Grassi2014}. Recently, \citet{Mackey2018} presented simulations with an updated NL99 chemistry network that incorporates updated reaction rates suggested by \citet{Gong2017}.
		
		Many of the simulations mentioned above have been performed using one of the two standard models of MC formation, namely the turbulent periodic box and the colliding flow simulations. The \textbf{Turbulent box (TB)} simulations became popular following the understanding that the energy dissipation in supersonic turbulence observed in MC conditions is comparable to the typical free-fall time \citep{MacLow1998,Stone98}, while the MCs themselves are a few free-fall times old; thus, energy has to be continuously injected into the MC to maintain such turbulence. In the last two decades, the nature of MCs as dynamical objects dominated by turbulence has been explored via simulations of driven supersonic turbulence \citep[e.g.][]{Padoan1999,MacLow04,Kritsuk2011,Federrath2013}. Recently, \citet{Padoan2016} suggested that such TB models are close approximations to the supernova-driven turbulence in the ISM. By now, multiple TB simulations using coupled chemistry have been performed \citep[e.g.][]{Glover07a,Glover10,Walch11,Micic12,Glover2012chemcompare,Padoan2017}. 
		
		The TB models describe random flows, and to form a dense cloud, some sort of convergent flow is required. The \textbf{Colliding flow (CF)} simulation is a simple scenario to model such gas accumulation, where two flows of warm neutral medium (WNM) converge to quickly form clumps of cold molecular gas in the resulting dense and turbulent collision interface. An extensive set of CF simulations have been performed \citep[e.g.][]{Paredes99,Heitsch2006,Semadeni2007,Hennebelle07a, Hennebelle07b, Hennebelle2008, Banerjee09,Clark2012collflow,Koertgen2015,Fogerty2016} in which various dynamical and thermal instabilities generate turbulence at the collision interface and the inflowing WNM from both sides injects energy into the system. These CF simulations are also able to reproduce the typical velocity dispersion and column densities observed in the MCs \citep[e.g.][]{Klessen10, Heitsch2011, Valdivia2016}.
	
	Some of these studies have discussed the influence of the numerical resolution required to model the structure and composition of MCs.
	\citet{Federrath2013} perform very high resolution isothermal TB simulations without gravity and concluded that a numerical resolution of 1024$^3$ or higher is needed to get a converged density distribution with $<20\%$ difference from the ``infinite-resolution limit". 
	\citet{Gong2018} conclude that a resolution of $\sim2$~pc is sufficient to obtain a converged CO-to-H$_2$ conversion factor in the chemical post-processing of their simulation of a region of a galactic disk; however, the convergence of the total CO abundance is not achieved for a spatial resolution of up to $\sim1$~pc. In the TB simulations by \citet{Glover10} and \citet{Micic12}, the CF simulations by \citet{Valdivia2016}, and the zoom-in simulations of MC evolution from the turbulent ISM by \citet{Seifried2017b} -- all of which have coupled chemistry networks  -- a spatial resolution of $\sim 0.1$~pc seems to be  necessary and also sufficient to follow the H$_2$ abundance accurately. CO has a comparatively complex chemistry, and its abundance is mostly dependent on the local density and shielding properties. Although many studies with a coupled chemistry network have done some convergence tests, the spatial resolution needed to model the different chemical species in such 3D simulations has not been consistently determined. Since the idea of dynamical modelling with coupled chemistry is flourishing, it is necessary to have a thorough look into the resolution requirements to model the chemical evolution of turbulent MCs. The multiple TB and CF simulations in this paper are designed to investigate these resolution requirements and will serve to indicate whether such requirements have been met in existing studies.
	
	The paper is organized as follows: the numerical methods implemented in the simulations are described in Section~\ref{sec:methods} and the setups of the TB and CF simulations along with the complete list of simulation runs are described in Section~\ref{sec:setup}. The results of the simulations are presented in Section~\ref{sec:results} and the resolution requirements for converged H$_2$ and CO production are then derived in Section~\ref{sec:H2_CO_res_criteria}. Finally, the conclusions are drawn in Section \ref{sec:conclusions} followed by some complementary information in the Appendix.


\section{Numerical Methods}
	\label{sec:methods}
	The three-dimensional (3D) MHD adaptive mesh refinement (AMR) code \textsc{Flash4} \citep{Fryxell00,Dubey08} is used to perform the TB and CF simulations. The included physical processes are introduced below \citep[see][ for more details]{Walch15_SILCC1}.
	
\subsection{Magnetohydrodynamics} \label{sec:mhd_eq}
We solve the ideal MHD equations using the five-wave Bouchut MHD solver \citep{Waagan2011}. Additional source terms for self-gravity, turbulent driving, and radiative heating and cooling are handled using operator splitting. Since this project studies the chemical evolution in hydrodynamic simulations, $\mathbf{B} = 0$ throughout this paper. The impact of magnetic fields will be discussed in a subsequent paper (Weis et al. in prep). The set of equations reads:
\begin{flalign}
&\frac{\upartial \rho}{\upartial t} + \nabla \cdot  (\rho \mathbf{v} ) = 0 \, ,\\
\label{eq:hydro_mom}
&\frac{\upartial (\rho \mathbf{v})}{\upartial t} + \nabla \cdot \left [ \rho \mathbf{v} \mathbf{v}^\text{T} + \left( P + \frac{\mathbf{B}^2}{8\upi} \mathbf{I} \right) - \frac{\mathbf{B} \mathbf{B}^\text{T}}{4 \upi} \right]  = \rho (\mathbf{g + F}) \, ,\\
\label{eq:hydro_ener}
&\frac{\upartial E}{\upartial t} + \nabla \cdot \left[ \left( E + \frac{\mathbf{B}^2}{8\upi} +\frac{P}{\rho} \right)\mathbf{v} - \frac{(\mathbf{B} \cdot\mathbf{v}) \mathbf{B}}{4\upi} \right] = \rho \mathbf{v} \cdot (\mathbf{g + F}) +\dot{u}_\text{chem} \, ,\\
&\frac{\upartial \mathbf{B}}{\upartial t} - \nabla \times (\mathbf{v} \times \mathbf{B}) = 0 \, ,\\
&E = u + \frac{\rho \mathbf{v}^2}{2} + \frac{\mathbf{B}^2}{8\upi} \, ,
\end{flalign}	
where $t$ is time, $\rho$ is the density,  $\mathbf{v}$ is the gas velocity, $P = (\gamma -1) u$ is the thermal pressure ($u$ is the internal energy density, $\gamma = 5/3$ is the adiabatic index of the ideal gas), $\mathbf{B}$ is the magnetic field, $\mathbf{I}$ is the identity matrix, $\mathbf{g}$ is the acceleration due to self-gravity, $\mathbf{F}$ is the random, turbulent acceleration introduced for the TB simulations ($\mathbf{F} = 0$ for the CF simulations), $E$ is the total energy density and $\dot{u}_\text{chem}$ is the net rate of change of internal energy due to the heating from the attenuated uniform background interstellar radiation field (ISRF) as well as due to radiative cooling. 

\subsection{Self-gravity} \label{sec:selfgrav}
The acceleration due to gas self-gravity $(\mathbf{g} = -\nabla \Phi )$ is obtained by solving the Poisson equation for the gas
	\begin{equation}
		\Delta\Phi = 4\pi G \rho  \, ,
	\end{equation}
where $\Phi$ and $G$ refer to the gravitational potential and the gravitational constant, respectively. The \textsc{Gravity} module by \citet[][]{Wuensch2018} calculates $\mathbf{g}$ with an OctTree-based method. We use the geometric Multipole Acceptance Criterion (MAC) introduced by \citet{Barnes1986} with a limiting angle of $\theta_\text{lim} = 0.5$. In this project, self-gravity is activated after 10~Myr for the TB simulations, i.e. after the turbulence has developed for $\sim 3$ crossing times. The CF simulations, on the other hand, incorporate self-gravity from the very beginning.
	 
\subsection{Dust shielding and molecular (self-)shielding}
\label{sec:opticaldepth}
	The attenuation of the ISRF due to dust, H$_2$ and CO (self-)shielding and the shielding of CO by H$_2$ in every single cell in the computational domain is incorporated using the \textsc{OpticalDepth} module \citep{Wuensch2018}. The algorithm is similar to the \textsc{Treecol} algorithm \citep{Clark12} and has been used in \citet[][]{Gatto2015,Walch15_SILCC1, Girichidis2016_SILCC2, Gatto2017_SILCC3,Peters17_SILCC4} and \citet{ Seifried2017b}. When constructing the column density and extinction maps for each cell, we use the recommended number of \textsc{Healpix} pixels of $N_\text{PIX} = 48$ corresponding to $\theta_\text{lim} = 0.5$ \citep[see][]{Wuensch2018}. Furthermore, only the contributions from gas within a sphere with radius 16~pc  from the cell of interest are included. This radius corresponds to half the box side length in the TB simulations and half of the y-- and z-- side length for the CF simulations and avoids duplicate contributions from the same gaseous sub-structure across periodic boundary conditions (see Section~\ref{sec:setup}).
	
	\subsection{Turbulence}
	\label{sec:turbulence}
	In the TB simulations, turbulent velocities are driven in the gas using a random forcing field, which is evolved in time via the Ornstein-Uhlenbeck process \citep{Konstandin2015}. In Fourier space, the forcing acceleration ($\mathbf{F}$) in Equations~\ref{eq:hydro_mom}  and \ref{eq:hydro_ener} evolves as
	\begin{equation}
		\text{d}\mathbf{\tilde{F}}(\mathbf{k},t) = \tilde{F_0}(\mathbf{k},t) \mathcal{P}^\zeta(\mathbf{k}) \frac{\text{d} \mathbf{W}(t)}{T_\text{ac}} - \mathbf{\tilde{F}}(\mathbf{k},t) \frac{\text{d}t}{T_\text{ac}} \, ,
	\end{equation}
	where $\mathbf{\tilde{F}}(\mathbf{k},t)$ is the Fourier transform of the forcing acceleration as a function of wave-vector $\mathbf{k}$ and time $t$, $\tilde{F_0}(\mathbf{k},t)$ is the forcing amplitude of the $\mathbf{k}$ mode, $\mathbf{W}(t)$ is the Wiener process that introduces Gaussian random perturbations in $\tilde{F_0}(\mathbf{k},t)$, $\mathcal{P}^\zeta(\mathbf{k})$ is the projection operator that projects the perturbed Fourier amplitudes to $\mathbf{k}$ in order to obtain a desired mixture of solenoidal to compressive modes (here we choose a thermal mix of modes), and $T_\text{ac}= L/v_\text{rms}$ is the auto-correlation time. The last term represents the exponential decay of the Fourier modes. Turbulence is introduced on the box scale such that  $\tilde{F_0}(\mathbf{k},t) \neq 0$ only for $\vert\mathbf{k}\vert = 1$, where $\vert\mathbf{k}\vert$ is measured in units of $2 \upi / L$, such that the largest eddy is formed on the box scale with $L = 32$~pc (see Section~\ref{sec:TB_setup}). The forcing mechanism is configured to maintain the mass-weighted 3D root-mean-square velocity $v_\text{rms}$ of the gas at 10~\kms, and to re-normalise the bulk velocity of the simulation domain to zero in every forcing step throughout the simulation time. Therefore, we set $T_\text{ac} = L/v_\text{rms} = 3.13$~Myr.
	
	\subsection{Chemical Model, Heating and Cooling}
	\label{sec:chemistry}
		The chemical network used in this project is the ``NL99" network \citep[see][for more details]{Glover2012chemcompare}, which combines the hydrogen chemistry taken from \citet[][]{Glover07a, Glover07b} and the CO  network from \citet{NL99}. The network tracks the abundances of 14 chemical species plus free electrons (non-equilibrium evolution of H$_2$, H$^+$, C$^+$, CH$_x$, OH$_x$, CO, HCO$^+$, He$^+$, M$^+$ while atomic H, He, C, O, M are calculated via conservation laws).  Here, CH$_x$ and OH$_x$ refer to intermediate carbon-bearing or oxygen-bearing species such as CH, CH$_{2}$, OH, H$_2$O or OH$^{+}$, and M$^+$ to ionized metals. This network considers multiple pathways of CO formation via CH$_x$ and OH$_x$ species, and CO destruction via photodissociation as well as via reaction with ions such as H$_3 ^+$ and He$^+$ that are very sensitive to the cosmic ray ionization rate \citep{Bisbas2015} and the X-ray energy density \citep{Mackey2018}.
		
		Cooling and heating due to chemical reactions, the photoelectric effect on dust grains, UV radiation,  X-ray, and cosmic rays via the diffuse ISRF with strength $G_0 = 1.7$ in units of the Habing field \citep{Habing68, Draine78} are taken into account. The relative abundances (number of atoms relative to total H nuclei) of total He, C, O, and M are $x_\text{He} = 0.1$, $x_\text{C} = 1.4 \times 10^{-4}$, $x_\text{O} = 3.2 \times 10^{-4}$, and $x_\text{M} = 1.0 \times 10^{-7}$, respectively \citep{Sembach2000,Szucs2014}. The cosmic ray ionization rate of atomic hydrogen is $\zeta_\text{H} = 3\times 10^{-17}$~s$^{-1}$ and the cosmic ray ionization rate for other chemical species are then scaled as $\zeta_{\text{H}_2} = 2 \times \zeta_\text{H}$,  $\zeta_\text{He} = 1.09 \times \zeta_\text{H}$, and $\zeta_\text{C} = 3.83 \times \zeta_\text{H}$ using the factors from \citet{Liszt2003} based on data from the UMIST\footnote{\url{http://www.rate99.co.uk}} database.
		

\section{Simulation Setup}
	\label{sec:setup}
	
	\subsection{Turbulent Box (TB) simulation}
		\label{sec:TB_setup}
\begin{figure}
		\includegraphics[width=\linewidth]{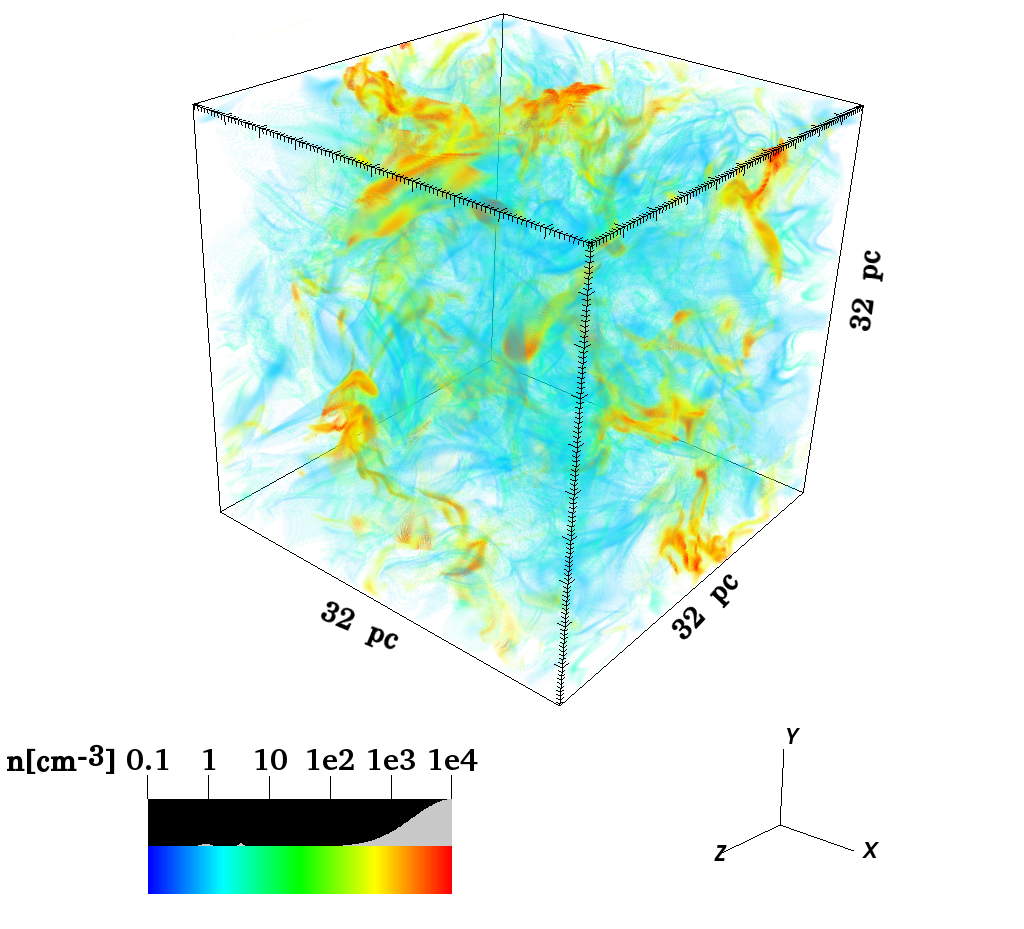}
	 	\caption{A snapshot of the TB-R6-n030 simulation showing the volume-rendered gas number density distribution  at $t = 20$~Myr. Here, the opacity of the surrounding low density medium is reduced to reveal the relatively dense structures formed inside, as indicated by the opacity bar above the color bar.}
	 	\label{fig:TB_snapshot}
	\end{figure}

		The TB simulations are performed in a (32~pc)$^3 $ box -- shown in Figure~\ref{fig:TB_snapshot} -- with periodic boundary conditions for hydrodynamics and gravity. Typically, turbulent box simulations start with uniform atomic gas: \citet{Glover07a} performed TB simulations with three different initial number densities $n_0 = 10,\; 30,$~and~100~\pcmc, \citet{Glover10} with $n_0=300$~\pcmc, \citet{Micic12} with $n_0=30$~and~300~\pcmc, \citet{Glover2012chemcompare} with $n_0=100$~and~300~\pcmc, and \citet{Padoan2017} simulated supernova-driven turbulence in a periodic box with $n_0=5$~\pcmc. The high density medium naturally forms molecular gas much faster than the low density medium. 
		
		The initial densities of H nuclei in the TB simulations presented here span two orders of magnitude: $n_0 = 3$~\pcmc (low density), $n_0 = 30$~\pcmc (intermediate density), and $n_0 = 300$~\pcmc (high density). Table~\ref{tab:sim_init_cond} shows the initial conditions, namely the equilibrium values of temperature, sound speed, and the abundance of the chemical species relative to H nuclei in a quiescent medium (without gravity, shielding, and turbulence) for each initial density considered in the simulations. The equilibrium conditions are calculated self-consistently by running the chemical network on a one-zone model with density $n_0$ for 10$^4$~Myr and assuming no shielding of the gas. The gas remains atomic with all of the hydrogen in H and all of the carbon atoms as C$^+$ ions.
		
The isothermal sound speed, $c_\text{s}$, in Table~\ref{tab:sim_init_cond} is calculated via
	\begin{equation}
		\label{eq:snd_spd}
		c_\text{s} = \sqrt{\frac{k_\text{B} T}{\mu \, m_\text{p}}}\, ,
	\end{equation}
	where $k_B$ is the Boltzmann constant, $T$ is the temperature of the gas in equilibrium, $ \mu=1.27$ is the mean molecular weight of the gas, and m$_\text{p}$ is the proton mass.
	
	\begin{table}
		\caption{Initial conditions for the different TB and CF simulations. Initial gas density $\rho_0$ (2$^{\rm nd}$ column), number density $n_0$ of H nuclei (3$^{\rm rd}$ columns), temperature $T_0$ (4$^{\rm th}$ column), and the sound speed $c_s$ (5$^{\rm th}$ column). The initial relative abundances of  $x_{\text{He}} = 0.1$, $x_{\text{C}} = 1.4 \times 10^{-4}$, $x_{\text{O}} = 3.2 \times 10^{-4}$, and $x_{\text{M}} = 1 \times 10^{-7}$ are common for all the TB and CF simulations. The initially atomic gas with $ \mu=1.27$ contains all of the hydrogen as H atoms and all of the carbon as C$^+$ ions.}
		\centering
		\begin{tabular}{l | l | l | l | l | l | l | l | l | l | l }
			\hline
			\hline
			Setup & $\rho_0$ [g~\pcmc] & $n_0$ [\pcmc] & $T_0$ [K] & $c_s$ [\kms] \\
			\hline
			TB & $7\times10^{-24}$ & 3      & 880 & 2.4 \\
			TB & $7\times10^{-23}$ & 30    & 122 & 0.9 \\
			TB & $7\times10^{-22}$ & 300  & 88   & 0.7 \\
			CF & $1.7\times10^{-24}$ & 0.75  & 4082 & 5.1 \\
			\hline
			\hline
		\end{tabular}
		\label{tab:sim_init_cond}
	\end{table}
		
	The simulations are named as  TB-R[x]-n[ddd], where TB denotes the Turbulent Box model, R[x] denotes the refinement level x that corresponds to a certain uniform resolution and n[ddd] denotes the initial number density of H nuclei in the homogeneous gas. The TB simulations are evolved with four different uniform resolutions ranging from the cell size of $\Delta x = 0.5$~pc ($64^3$; R4) to $\Delta x = 0.063$~pc ($512^3$; R7). For example, TB-R7-n030 is a turbulent box simulation with $n_0=30$~\pcmc and refinement level 7 (i.e. $512^3$). Note that the TB simulations do not use the AMR capabilities of \textsc{Flash}; however, the notation of the refinement level (e.g. R4) is used to be consistent with the nomenclature for the CF simulations. Supersonic turbulence is constantly driven in the simulation box to maintain $v_\text{rms} = 10$~\kms that results in the turbulent crossing time $T_\text{ac} = 3.13$~Myr (see Section~\ref{sec:turbulence}). TB simulations with lower $v_\text{rms}$ produce lower molecular content \citep[][]{Micic12}. To allow for a direct comparison of the different resolution runs, the same turbulence driving pattern is used in all the simulations with the (32~pc)$^3$ box. 
	
	In addition, another set of test simulations with $n_0=30$~\pcmc and  $v_\text{rms} = 1,2,3,6$~\kms are performed in a smaller (8~pc)$^3 $ box. These simulations have the uniform spatial resolution corresponding to the TB-R[5,6,7]-n030 runs, and are designed to test the effect of $v_\text{rms}$ on the convergence of the H$_2$ chemistry (only used in Section~\ref{sec:TB_res_req}). These 12 special simulations have a ``-v[x]" tag at the end, which indicates the driving velocity, and are run only up to 10~Myr. 
	
	All these hydrodynamic simulations exclude magnetic fields and self-gravity is activated for the simulations only after 10~Myr, when the turbulence is well-developed. The dense MC regions form and evolve in the TB simulations up to 20~Myr. The runs with $n_0=300$~\pcmc are stopped shortly after 10~Myr because the medium is so dense that some regions collapse immediately due to gravity (further explanation in Section~\ref{sec:TB_CF_gen_diff}). The complete list of the TB simulations is given in Table~\ref{tab:runs}.
	
	\subsection{Colliding Flow (CF) simulation}
		\label{sec:CF_setup}
\begin{figure}
				\includegraphics[width=\linewidth]{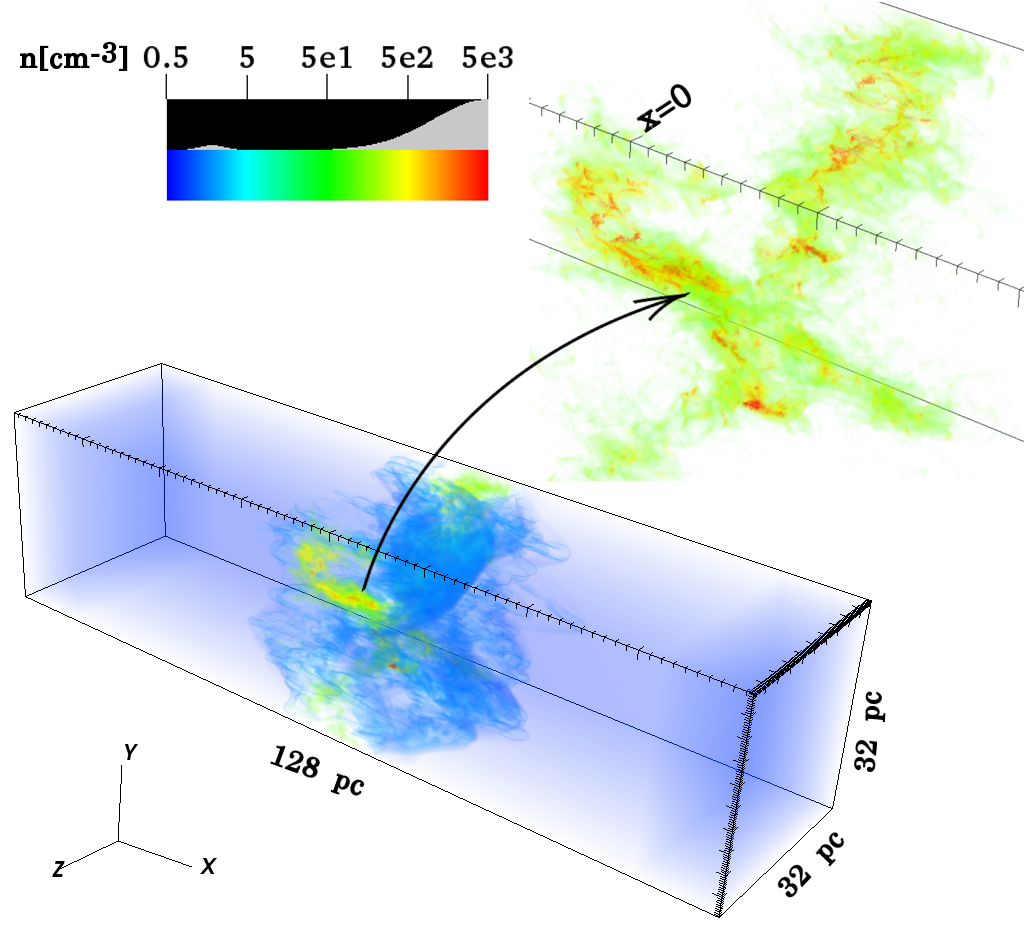}
			 	\caption{A snapshot of the CF-R6 simulation, showing the volume-rendered gas number density distribution at $t = 20$~Myr. As in Figure~\ref{fig:TB_snapshot}, the opacity of the incoming low density is reduced to show the dense objects formed in the centre. The zoomed-in region shows the formed filamentary sub-structure.}
			 	\label{fig:CF_snapshot}
			\end{figure}

			The CF simulations by \citet[][]{Heitsch2006,Semadeni2007,Clark2012collflow,Koertgen2015} and \citet{Valdivia2016} follow the collision of two streams of WNM with density $n_0 = 1$~\pcmc, the flow velocity ranging from 5~\kms to 31~\kms, and the domain size ranging from 44~pc to 256~pc. The domain of the CF simulation presented here is a 128~pc $\times$ 32pc $\times$ 32~pc rectangular cuboid with inflow boundary conditions in the $x$--direction and periodic boundaries in the $y$-- and $z$--directions  (shown in Figure~\ref{fig:CF_snapshot}). Similarly, the gravity boundary condition is isolated in the $x$--direction and periodic along the $y$-- and $z$--directions \citep[see][for the implementation of mixed gravity boundary conditions]{Wuensch2018}. The whole domain is initially filled with a warm, uniform density medium with $n_0 = 0.75$~\pcmc. As in section~\ref{sec:TB_setup}, we use the chemical network to calculate the equilibrium temperature and chemical composition for this $n_0$. We summarize the initial conditions in Table~\ref{tab:sim_init_cond}. 
			
\begin{figure}
				\includegraphics[width=\columnwidth]{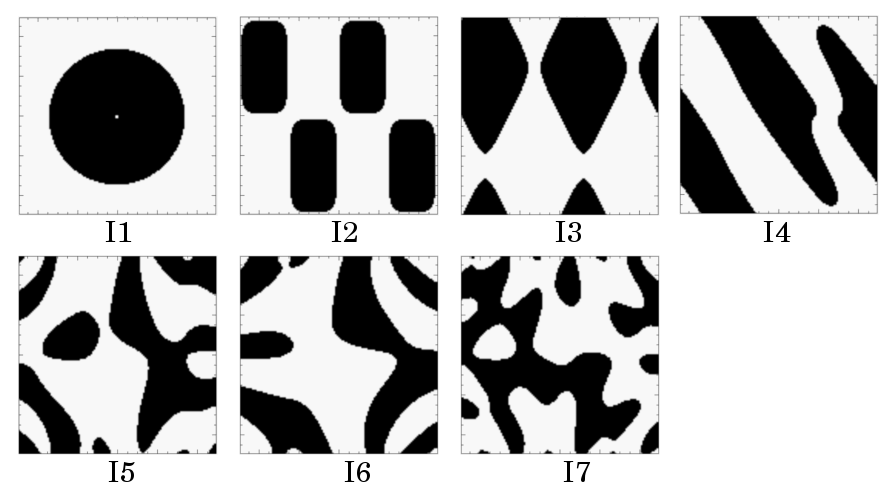}
				\caption{ Slices of various collision interfaces at the $x = 0$ plane, showing the $x$-component of the velocity. Black regions denote gas flowing out of the plane of paper (with positive $v_x$) and white regions have negative $v_x$. All the colliding flow simulations presented in this paper have the interface I5.} 
				\label{fig:CF_interface_pictures}
			\end{figure}

\begin{figure}
				\includegraphics[width=\columnwidth]{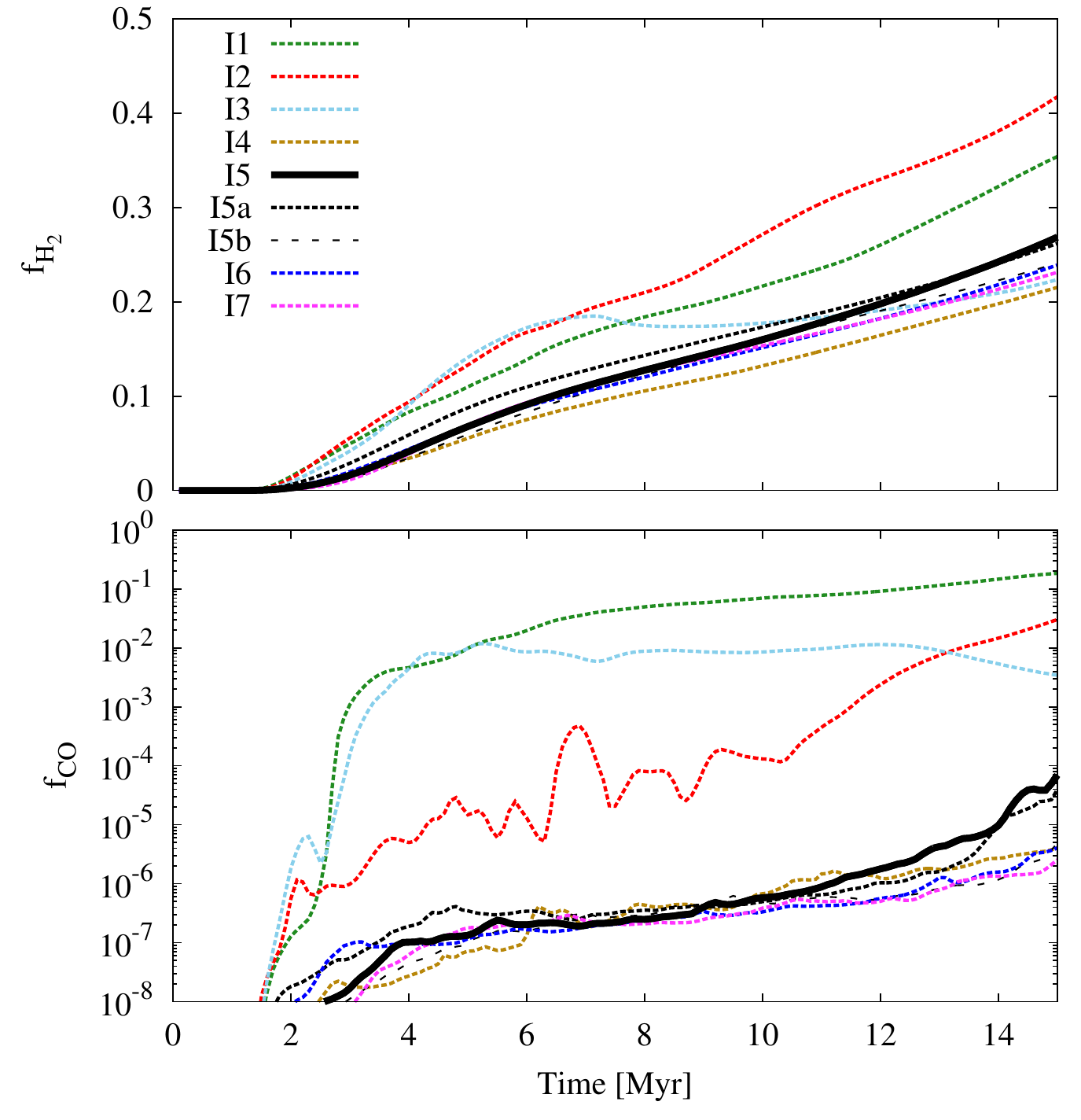}
				\caption{ Time evolution of the mass fraction of H$_2$ (top) and CO (bottom) for the different interfaces shown in Figure~\ref{fig:CF_interface_pictures}. The mass fractions $f_{\text{H}_2}$ and $f_{\text{CO}}$ are defined in Equations~\ref{eq:fH2} and \ref{eq:fCO}, respectively. The interface I5 results in an intermediate amount of molecular gas and is chosen for all other CF simulations in this paper.} 
				\label{fig:CF_interface_H2CO}
			\end{figure}

			An irregular interface is constructed at the centre of the simulation box to trigger nonlinear thin-shell instabilities. This is an extension of the one-dimensional sinusoidal interface implementation in \citet{Heitsch2006}. The collision interface that separates the flows is created by perturbing the $x = 0$ plane using the function
			\begin{equation}
			\label{eq:interface_eq}
				x' = A \left[ \cos (2-yz) \cos (k_y y+\phi) + \cos (0.5- yz) \sin (k_z z) \right]
			\end{equation}
			where $x'$ is the $x$--coordinate of the perturbed interface. The parameters that allow one to vary the structure of the collision interface are the amplitude of the perturbation $A$, the wave-numbers along the $y$-- and $z$--axis, $k_y$ and $k_z$, and the phase $\phi $. While the velocities in $y$-- and $z$--direction are initially zero, the $x$--velocity on either side of this interface is assigned with a value of $v_x = \pm13.6$~\kms, so that the collision occurs immediately upon the start of the simulation.
			Figure~\ref{fig:CF_interface_pictures} shows the slices of multiple colliding interfaces at the $x = 0$  plane (black regions correspond to positive $v_x$, white regions to negative $v_x$); Figure~\ref{fig:CF_interface_H2CO} shows the H$_2$ and CO content developing in the CF simulations up to 15 Myr for 9 different interfaces. The interfaces with large, homogeneous sub-structures (I1, I2 and I3) do not provide a lot of surface area for the opposite flows to interact. As a result, gas is easily collected in homogeneous ``pockets" and much more H$_2$ and CO is produced due to more effective shielding. In all other interfaces that lead to more sub-structure, the molecular content is systematically smaller. Interfaces I5a, I5, and I5b are obtained by changing the amplitude ($A = $ 1~pc, 1.6~pc, and 2~pc, respectively) in Eq.~\ref{eq:interface_eq}, while all other parameters are fixed; the slice of the collision interface at the $x = 0$ plane in Figure~\ref{fig:CF_interface_pictures} is identical for all three cases. Figure~\ref{fig:CF_interface_H2CO} shows that the molecular content changes slightly with the change in $A$. For the presented CF simulations, the interface I5 with $A =$ 1.6~pc, $k_y = 2$, $k_z = 1$, and  $\phi = 0$, that produces an intermediate amount of molecular gas, is chosen. The system is evolved up to 20~Myr to allow for dense MC cores to form and evolve.
				
			The complete list of CF simulations is given in Table~\ref{tab:runs}. The simulations are named CF-R[x] where CF denotes the Colliding Flow model and R[x] denotes the refinement level x that corresponds to an effective resolution via the AMR technique. The low-resolution CF-R4 and CF-R5 runs have uniform resolution. For all higher resolution runs, the base-grid resolution is $\Delta x = 0.25$~pc (corresponding to R5) and the effective resolution is achieved using  AMR. For refinement level 6 ($\Delta x = 0.125$~pc), we refine on a threshold on the second spatial derivative of the gas density (\textsc{Paramesh} refinement). For higher refinement levels, we use the Jeans criterion \citep{Truelove1997}, which requires that the Jeans length,
			\begin{equation}
				\label{eq:jeans_length}
				\lambda_\text{J} = \sqrt{\frac{\pi c_\text{s} ^2}{\rho G}} \, ,
			\end{equation}
			 is resolved with at least N$_\text{Jeans}$ cells in one dimension. Here we use N$_\text{Jeans}=8$.
			
			The CF simulation domain is constructed using 4$\times$(32~pc)$^3$ boxes (i.e. blocks) stacked along the $x$--direction, such that the first refinement level (R = 1) contains four blocks, each with $8^3$ cells. For each higher refinement level, the number of blocks along each dimension increases by 2; thus, the cell size obtained for refinement level R is
			\begin{equation}
				\Delta x = \frac{32\text{ pc}}{2^{R-1} \times 8} = \frac{32\text{ pc}}{2 ^{\text{R}+2} } = 2 ^{3 - \text{R}}\text{ pc} \, .
			\end{equation}
					
			\begin{table}
				\caption{List of simulations performed for this study. TB indicates the Turbulent Box setup and CF indicates the Colliding Flows setup. R[4-10] denotes the refinement level, where R4 corresponds to a cell size of  $\Delta x = 0.500$~pc and R10 corresponds to an effective resolution of $\Delta x = 0.008$~pc. There are two sets of TB runs, one in a (32 pc)$^3$ box with $v_{\rm rms}=10$~\kms and one in a (8 pc)$^3$ box with different  $v_{\rm rms}$. The latter has a ``-v[x]" tag and is used in Section~\ref{sec:TB_res_req} to verify our resolution requirement. In the TB runs, the initial density of the H nuclei in the homogeneous medium is denoted by the ending tag (e.g. n003 corresponds to  $n_0 = 3$~\pcmc, see column 3). The (effective) resolution of the simulations is denoted by $\Delta x$. In the TB runs, the value in parentheses denotes the number of finite volume cells used. In the CF runs, the ``AMR" in parentheses denotes that the effective resolution is obtained via adaptive mesh refinement.}
				\centering
				\begin{tabular}{ l | l | l | l }
					\hline
					\hline
					Run & $\Delta x$ & $n_0$& $v_\text{rms}$\\
					 &  &[\pcmc] & [\kms]\\
					\hline
					\multicolumn{4}{c}{Turbulent box: box size (32 pc)$^3$} \\
					\hline
					TB-R4-n003  & 0.500~pc (64$^3$)  & 3   & 10\\
					TB-R4-n030  & 0.500~pc           & 30  & 10\\
					TB-R4-n300  & 0.500~pc           & 300 & 10\\
					TB-R5-n003  & 0.250~pc (128$^3$) & 3  & 10\\
					TB-R5-n030  & 0.250~pc           & 30  & 10\\
					TB-R5-n300  & 0.250~pc           & 300 & 10\\
					TB-R6-n003  & 0.125~pc (256$^3$) & 3  & 10 \\
					TB-R6-n030  & 0.125~pc           & 30  & 10\\
					TB-R6-n300  & 0.125~pc           & 300 & 10\\
					TB-R7-n003  & 0.063~pc (512$^3$) & 3   & 10\\
					TB-R7-n030  & 0.063~pc           & 30  & 10\\
					TB-R7-n300  & 0.063~pc           & 300  & 10\\
					\hline
					\multicolumn{4}{c}{Small turbulent box: box size (8 pc)$^3$}\\
					\hline
					TB-R5-n030-v1  & 0.250~pc (32$^3$) & 30  & 1\\
					TB-R5-n030-v2  & 0.250~pc           & 30  & 2\\
					TB-R5-n030-v3  & 0.250~pc           & 30 & 3\\
					TB-R5-n030-v6  & 0.250~pc           & 30 & 6\\
					TB-R6-n030-v1  & 0.125~pc (64$^3$) & 30  & 1\\
					TB-R6-n030-v2  & 0.125~pc           & 30  & 2\\
					TB-R6-n030-v3  & 0.125~pc           & 30 & 3\\
					TB-R6-n030-v6  & 0.125~pc           & 30 & 6\\
					TB-R7-n030-v1  & 0.063~pc (128$^3$) & 30  & 1\\
					TB-R7-n030-v2  & 0.063~pc           & 30  & 2\\
					TB-R7-n030-v3  & 0.063~pc           & 30 & 3\\
					TB-R7-n030-v6  & 0.063~pc           & 30 & 6\\
					\hline
					\multicolumn{4}{c}{Colliding flow} \\
					\hline
					CF-R4  & 0.500~pc & 0.75 & --\\
					CF-R5  & 0.250~pc & 0.75 & --\\
					CF-R6  & 0.125~pc (AMR) & 0.75 & --\\
					CF-R7  & 0.063~pc (AMR) & 0.75 & --\\
					CF-R8  & 0.032~pc (AMR) & 0.75 & --\\
					CF-R9  & 0.016~pc (AMR) & 0.75 & --\\
					CF-R10  & 0.008~pc (AMR) & 0.75 & --\\
					\hline
					\hline
				\end{tabular}
				\label{tab:runs}
			\end{table}
	

\section{Results}
\label{sec:results}
	The focus of this paper is to study the changes in the chemical composition in both the TB and CF models with increasing resolution. At first, the basic differences between the results of the TB and CF runs are outlined. Then, the  gas distribution as well as the formation of H$_2$ and CO in models with increasing resolution is discussed. In all of the presented results of the CF runs, only the gas with temperature T $<$ 4000~K is taken into account so that the inflowing warm gas is excluded.
	
	\subsection{Turbulent Box and Colliding Flows : General differences}
		\label{sec:TB_CF_gen_diff}
		The setups of the TB and CF simulations are quite different and thus, the gas evolves differently in each model. The turbulence in the TB runs is continuously driven and energy is injected to maintain a constant velocity dispersion. The turbulence in the CF simulations is a result of various instabilities that have been described in \citet{Heitsch2006} and the energy is injected via the incoming flows. Since the evolution of the gas distribution differs for each initial condition in the different TB and CF runs, the chemical evolution should also be different.
			
\begin{figure*}
				\includegraphics[width=\linewidth]{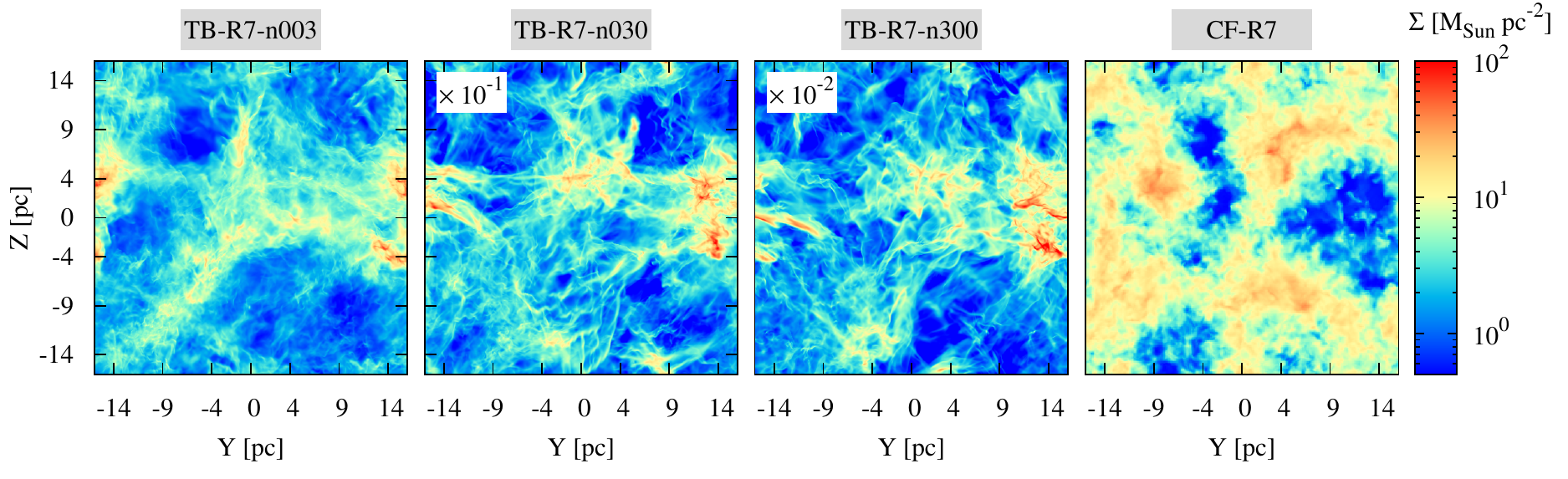}
				\caption{From left to right: Column density of the total gas in the TB-R7-n003, TB-R7-n030, TB-R7-n300 and CF-R7 simulations with $\Delta x = 0.063$~pc at $t = 10$~Myr, integrated along the $x$--direction. In order to compare the formed sub-structure for all simulations, the column densities of TB-R7-n030 and TB-R7-n300 runs are scaled with a factor of 0.1 and 0.01, respectively. For the CF-R7 run, only the gas near the collision layer with T $<$ 4000~K is taken into account.}
				\label{fig:TB_CF_coldens}
			\end{figure*}

			To get a general overview of the different gas morphology produced by the TB and CF runs, their column density maps are compared in Figure~\ref{fig:TB_CF_coldens}. The first three maps correspond to the TB simulations with various mean initial densities and the last map corresponds to the CF simulation, all at the same (effective) resolution ($\Delta x = 0.063$~pc, R7) and at $t = 10$~Myr. The TB-R7-n003, TB-R7-n030 and TB-R7-n300 runs produce gas column densities as high as $10^2, 10^3$ and $10^4$~M$_\odot $pc$^{-2}$, respectively (note the corresponding scaling shown in the figure). Driven turbulence creates filamentary structures in all TB simulations, that evolve until $t = 10$~Myr without self-gravity. On the other hand, the CF simulation with gravity produces comparatively larger, diffuse clumps with dense gas at their centre where the gas column density reaches $\Sigma\sim 10^2$~M$_\odot $pc$^{-2}$.
			
\begin{figure}
				\includegraphics[width=\linewidth]{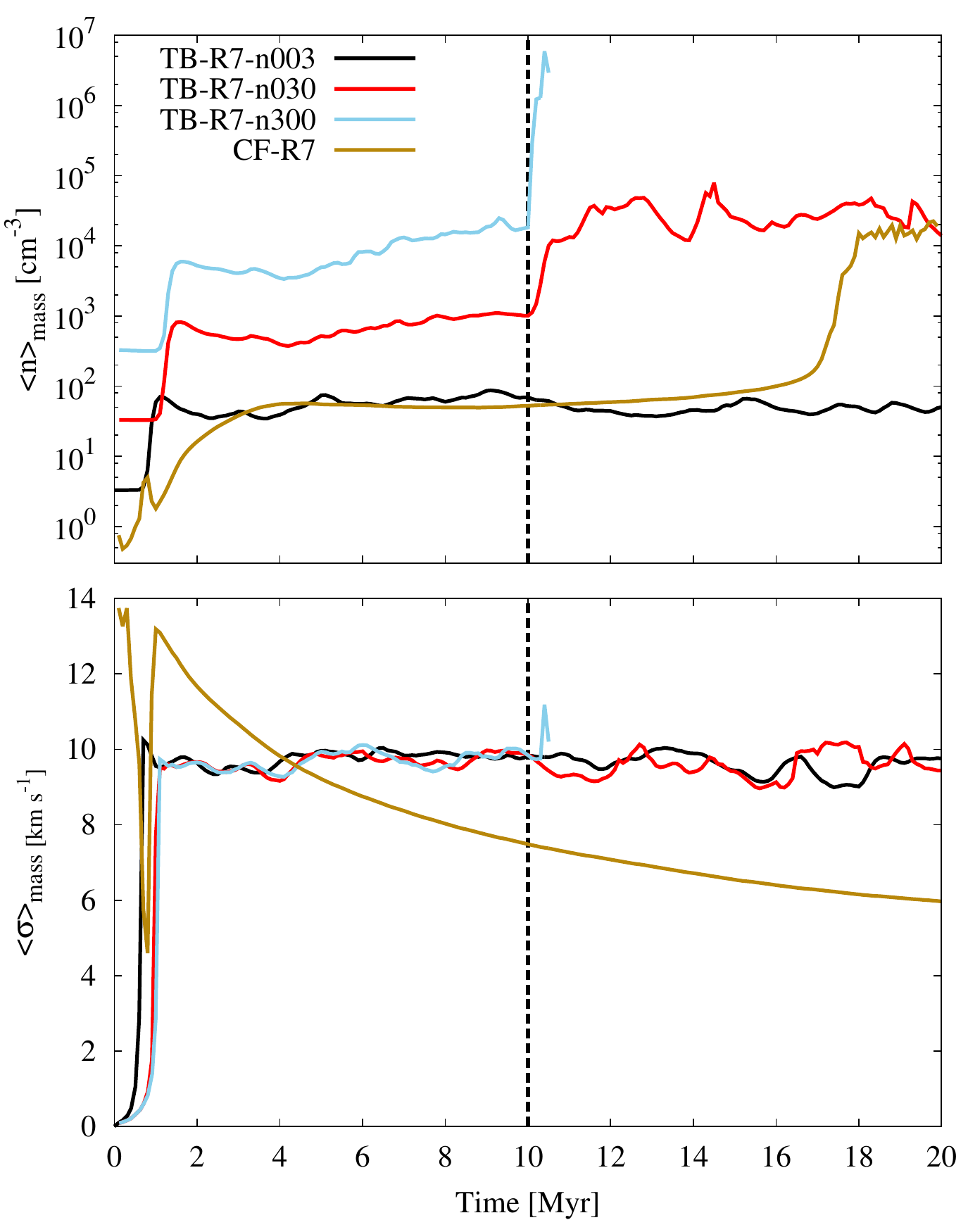}
				\caption{ Evolution of the mass-weighted mean number density $\langle n \rangle_\text{mass}$ (top) and  3D velocity dispersion $\langle \sigma \rangle_\text{mass}$ (bottom) of the gas in the TB-R7-[n003,n030,n300] and CF-R7 runs with $\Delta x = 0.063$~pc. The vertical, dashed line denotes when gravity is turned on in the TB simulations. The TB-R7-n300 run is stopped shortly after gravity is turned on because the box is dominated by free-fall collapse. For the CF-R7 run, only the gas near the collision interface with T $<$ 4000 K is taken into account.}
				\label{fig:TB_CF_dens_veldisp_evol}
			\end{figure}

			In Figure~\ref{fig:TB_CF_dens_veldisp_evol}, the evolution of the mass-weighted mean number density ($\langle n \rangle_\text{mass}$, top panel) and 3D velocity dispersion ($\langle \sigma \rangle_\text{mass}$, bottom panel) of the gas is shown for these four runs (TB-R7-n003, TB-R7-n030, TB-R7-n300 and CF-R7). The vertical, dashed line at $t = 10$~Myr denotes that gravity is switched on in the TB simulations after the turbulence is well-developed in the gas for $\sim 3$ crossing times. In all three TB runs, the jumps in $\langle n \rangle_\text{mass}$ after 1~Myr arise from local density enhancements when the homogeneous gas first experiences shocks due to the turbulent driving. The value of $\langle n \rangle_\text{mass}$ increases by an order of magnitude, and is maintained at this level until gas self-gravity is activated at 10~Myr. In the TB-R7-n003 run, the effect of gravity is not significant. In TB-R7-n030, gravity causes the fragmentation of gas and leads to high density regions. The results of the TB-R[4-7]-n300 runs with gravity are not used for analysis since, shortly after self-gravity is switched on at 10~Myr, most of the mass in these runs (above 30\%) is contained in unresolved dense structures that undergo gravitational collapse. In the CF-R7 run,  $\langle n \rangle_\text{mass}$ experiences a jump at $\sim$ 1~Myr when the colliding flow starts to  cool down in the centre. After that, it smoothly increases and remains at $\langle n \rangle_\text{mass} \sim 30$~\pcmc for the first 16~Myr, similar to the low density TB run. Self-gravity dominates the gas dynamics after $t \sim 16$~Myr and increases the mean gas density to $\langle n \rangle_\text{mass} \sim 2 \times 10^4$~\pcmc, similar to the intermediate density TB run. 
			
			The evolution of $\langle \sigma \rangle_\text{mass}$ in the bottom panel shows that all the TB runs maintain  $\langle \sigma \rangle_\text{mass} \approx 10$~\kms once the density inhomogeneities arise after 1~Myr. This is enforced by the turbulent driving described in Section~\ref{sec:turbulence}. On the other hand, for the CF run, the initially high velocity dispersion due to the colliding and cooling gas is gradually lowered to $\langle \sigma \rangle_\text{mass} = 6$~\kms as the denser regions begin to dominate the gas dynamics (see also Appendix~\ref{sec:appendix_veldisp}).
			
\subsection{Turbulent Box results}
\begin{figure*}
			\includegraphics[width=0.98\linewidth]{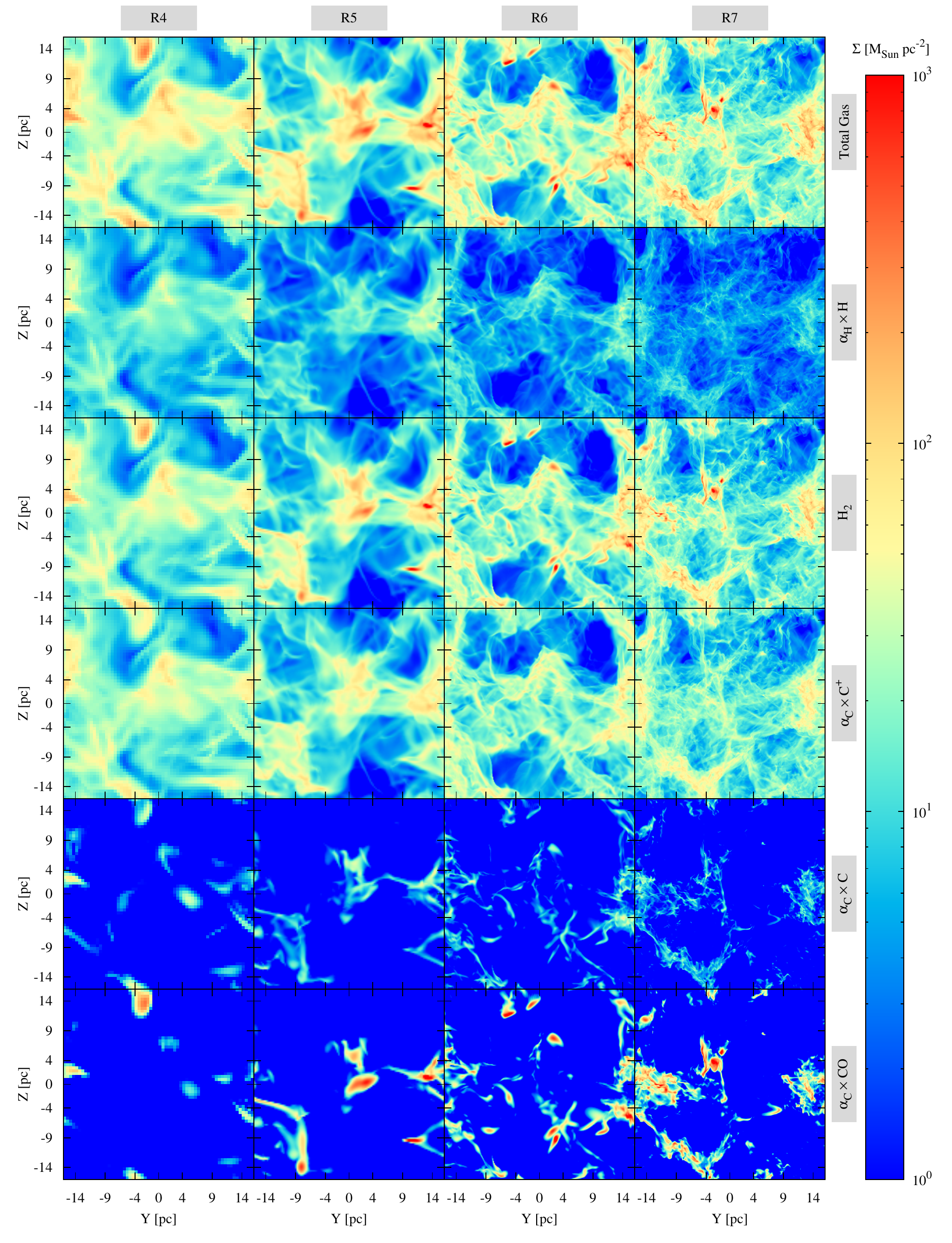}
			\caption{Column density integrated along the $x$--direction of the total gas, H, H$_2$, C$^+$, C and CO (top to bottom) for increasing resolution (left to right) in the TB-R[4-7]-n030 runs at $t = 20$~Myr. In order to make the column densities visible on the same color scale, the H column densities are scaled by a factor of $\alpha_\text{H} = 10$, and the C$^+$, C and CO column densities are scaled by a factor of $\alpha_\text{C} = 595$ (see Equation~\ref{eq:carbon_factor}). Higher resolution runs produce more filamentary structures with more H$_2$ and CO while the mass in H and C is reduced.}
			\label{fig:TB_n30_coldens}
		\end{figure*}

Figure~\ref{fig:TB_n30_coldens} displays the column densities integrated along the $x$--axis of the total gas, H, H$_2$, C$^+$, C and CO (from top to bottom) for increasing resolution (from left to right) in the TB-R[4-7]-n030 simulations at $t = 20$~Myr. The gas has already evolved with gravity for 10~Myr at this point. In order to make the low column density visible, the H column densities are scaled by $\alpha_\text{H} = 10$. Similarly, the column densities of the carbon species are scaled by a factor 
			\begin{equation}
				\label{eq:carbon_factor}
				\alpha_\text{C} = \frac{1}{x_\text{C,tot} \times \mu_\text{C} } = 595 \, , 
			\end{equation}
			where $x_\text{C,tot} = 1.4 \times 10^{-4}$ is the relative abundance of carbon atoms in the simulations and $\mu_\text{C} = m_\text{C}/m_\text{H} = 12.011$ is the atomic mass of carbon.
			
			Figure~\ref{fig:TB_n30_coldens} shows that both the large scale and the small scale gas morphology changes with resolution. Numerous denser but thinner and more filamentary regions form at higher resolution. The H column density decreases as the resolution increases since more H is converted to H$_2$ that resides either in the dense cores or in the filamentary structures. A low and diffuse C$^+$ column density is found throughout the simulation domain. The column density of C is also low and mainly traces the outskirts of the CO rich regions, showing regions of active photodissociation. In regions of  high CO column density, which are only found inside the H$_2$-rich regions, the amount of both C$^+$ and C is negligible. Thus, the distribution of all species greatly differ with increasing resolution. 
			
\subsubsection{Convergence study of H$_2$} \label{sec:TB_H2_conv_study}
In this section, the evolution of H$_2$ in the TB simulations with various initial conditions and resolutions are presented. Three different quantities are calculated to investigate the convergence of H$_2$ in the simulations. The mass-fraction of H$_2$ is defined as
			\begin{equation}
				\label{eq:fH2}
				f_{\text{H}_2} =  \frac{M_{\text{H}_2}}{M_\text{H,tot}} \equiv \frac{2 \, N_{\text{H}_2}}{N_\text{H,tot}}\, ,
			\end{equation}
			where $M_{\text{H}_2}$ is the total mass of H$_2$ and $M_\text{H,tot}$ is the total mass of all hydrogen atoms. The second, equivalent expression is the ratio of the number of H atoms in H$_2$ molecules ($2 \, N_{\text{H}_2}$) and total hydrogen atoms ($N_\text{H,tot}$). 
			
			The time-scale of H$_2$ formation in the simulation is defined as
			\begin{equation}
				\label{eq:tauH2form}
				\tau_{\text{H}_2, \text{form}} = \frac{M_\text{H,tot}}{\dot{M}_{\text{H}_2}} \, ,
			\end{equation}
			where $\dot{M}_{\text{H}_2}$ is the H$_2$ formation rate that is averaged over 2~Myr for every point. 
			
			Finally, taking the simulation with the highest refinement level, R$_\text{max}$, as a reference, the results for various resolutions are compared using the ratio
			\begin{equation}
				\label{eq:etaH2}
				\eta_{\text{H}_2} = \frac{f_{\text{H}_2}}{f_{\text{H}_2,\text{R}_\text{max}}} \, .
			\end{equation}
			
\begin{figure*}
				\includegraphics[width=\linewidth]{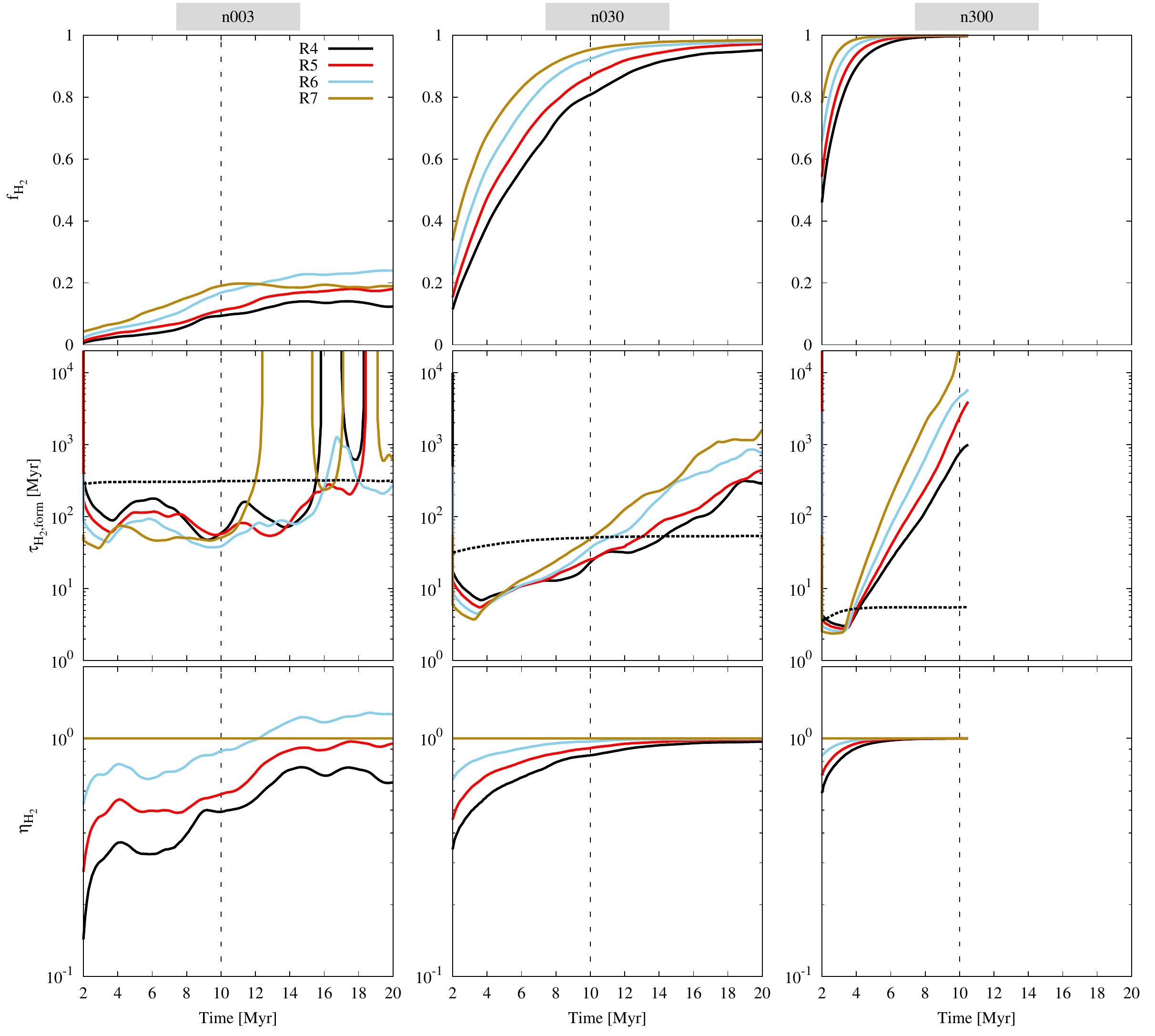}
				\caption{Evolution of H$_2$ in TB simulations with different resolutions. The vertical, dashed lines separate the time evolution without (t $< 10$~Myr) and with (t $> 10$~Myr) self-gravity. Results of different initial conditions are arranged in different columns. The high density runs in the third column are not evolved for long once self-gravity is turned on (see Figure~\ref{fig:TB_CF_dens_veldisp_evol}). First row: the evolution of the H$_2$ mass fraction ($f_{\text{H}_2}$, Equation~\ref{eq:fH2}). Second row: the evolution of the H$_2$ formation time-scale ($\tau_{\text{H}_2, \text{form}}$, Equation~\ref{eq:tauH2form}). When H$_2$ is being destroyed, $\tau_{\text{H}_2, \text{form}} \rightarrow \infty$. The dashed line is the mass-weighted mean equilibrium H$_2$ formation time-scale ($\langle \tau_{\text{H}_2,\text{eq}} \rangle_\text{mass}$, Equation~\ref{eq:H2_formation}). Third row: the evolution of $\eta_{\text{H}_2}$ (Equation~\ref{eq:etaH2}) to compare the results for various resolutions, taking the highest resolution run (R7) as a reference. The runs clearly show that the formation of H$_2$ is not converged in any of the three density regimes.}
				\label{fig:TB_H2_analysis}
			\end{figure*}

			Figure~\ref{fig:TB_H2_analysis} shows the results of low, intermediate, and high density runs (denoted by n003, n030, and n300) in the first, second, and third column, respectively. Since the first density inhomogeneities due to the turbulence appear after 1~Myr (see Figure~\ref{fig:TB_CF_dens_veldisp_evol}), the evolution of the molecular content starting from 2~Myr is shown. The vertical, dashed lines denote that gravity is switched on at $t=10$~Myr. Note that the TB-R[4-7]-n300 runs evolve only for a short time after 10~Myr due to the reasons mentioned in Section~\ref{sec:TB_CF_gen_diff}. The first row shows the evolution of $f_{\text{H}_2}$, with a maximum value of 1 when 100\% of the H atoms are in the form of H$_2$. In all three setups with different initial densities, the higher resolution runs generally have higher $f_{\text{H}_2}$ at any given time. One exception to this is the TB-R6-n003 run, that contains more H$_2$ than the TB-R7-n003 run after 10~Myr. The runs with low and intermediate density do not show any significant gravitational effects on the evolution of $f_{\text{H}_2}$ for any resolution. In the high density runs, the values of $f_{\text{H}_2}$ for various resolutions look similar towards the end of the simulations. This is trivial because the molecular content in these high initial density runs saturates and the gas becomes fully molecular regardless of the resolution. The large variation in the evolution of $f_{\text{H}_2}$ for different resolutions derives directly from the changing gas morphology, as shown in Figure \ref{fig:TB_n30_coldens}.
			 
			The second row shows the evolution of  $\tau_{\text{H}_2, \text{form}}$ for various resolutions. The dashed lines in each plot denote the mass-weighted mean value of the theoretical estimate for the equilibrium H$_2$ formation time-scale $\langle \tau_{\text{H}_2,\text{eq}} \rangle_\text{mass}$, which is given by \citet[]{Hollenbach1971,Hollenbach1989,Goldsmith2005}
			\begin{equation}
			\label{eq:H2_formation}
				\tau_{\text{H}_2,\text{eq}} \sim 10^3 \text{ Myr} \, \left( \frac{\bar{n}}{1 \text{ cm}^{-3}} \right)^{-1} \, ,
			\end{equation}
			where $\bar{n}$ is the mean gas number density in the region of interest. In all of the low, intermediate, and high density runs, the H$_2$ formation at all resolutions is faster than $\langle \tau_{\text{H}_2,\text{eq}} \rangle_\text{mass}$ before the saturation level is reached. This ``faster-than-average" H$_2$ formation in turbulent media has been shown by \citet{Glover07b} and recently noted by \citet{Micic12}, \citet{Valdivia2016}, and \citet{Seifried2017b}. For the low density runs in the first column, $\tau_{\text{H}_2, \text{form}} \rightarrow \infty$ when H$_2$ is being destroyed. There is a common trend for the runs in all three columns: after the first density inhomogeneities set in and H$_2$ formation starts, $\tau_{\text{H}_2, \text{form}}$ becomes shorter with increasing resolution. Consequently, the saturation level of $f_{\text{H}_2}$ is reached faster for higher resolution runs. After $f_{\text{H}_2}$ has reached saturation in the higher resolution runs, $\tau_{\text{H}_2, \text{form}}$ becomes longer for higher resolution. This is less obvious for the low density runs, but clearly visible in the intermediate and high density runs. 
			
			The third row shows the time evolution of $\eta_{\text{H}_2}$. In case of the TB runs, R$_\text{max}=$R7 with $\Delta x=0.063$~pc. Note that $\eta_{\text{H}_2}$ only reflects the trend of the simulations with increasing resolution, and $\eta_{\text{H}_2} = 1$ is not the expected converged state. In case of the low density runs, the TB-R4-n003 and TB-R5-n003 runs produce less H$_2$ than the TB-R7-n003 run ($\eta_{\text{H}_2} < 1$), while the TB-R6-n003 run has $\eta_{\text{H}_2} > 1$ shortly after 10~Myr.  In the intermediate and high density runs, $\eta_{\text{H}_2} < 1$ for all the lower resolution runs and $\eta_{\text{H}_2} \rightarrow 1$ faster with increasing resolution. Hence, although $\eta_{\text{H}_2} \rightarrow 1$ with time in all cases, the evolution of total H$_2$ is significantly different for the various resolutions and the formation of H$_2$ in runs with resolutions $\Delta x \geq 0.125$~pc is not converged. Since these simulations are computationally expensive ($\sim500,000$ CPU-hours for the TB-R7-n030 run), it is currently not possible to go to resolutions as high as achieved in the isothermal TB simulations by, for example, \citet{Federrath2013}.
			
\subsubsection{Convergence study of CO} \label{sec:TB_CO_conv_study}
Similar to the study for H$_2$, three quantities are calculated to study the formation of CO. The mass fraction of CO is defined as
			\begin{equation}
				\label{eq:fCO}
				f_{\text{CO}} =  \frac{\frac{12}{28} M_{\text{CO}}}{M_\text{C,tot}} \equiv \frac{N_\text{CO}}{N_\text{C,tot}} \, ,
			\end{equation}
			where $\frac{12}{28} M_{\text{CO}}$ is the total mass of carbon in CO molecules and $M_\text{C,tot}$ is the total mass of all the carbon atoms. The equivalent expression is the ratio of the number of carbon atoms in CO molecules, $N_\text{CO}$, to total carbon atoms, $N_\text{C,tot}$. 
			
The CO formation time-scale in the simulation is defined as
			\begin{equation}
				\label{eq:tauCOform}
				\tau_{\text{CO}, \text{form}} = \frac{M_\text{C,tot}}{\frac{12}{28}  \dot{M}_{\text{CO}}} \, ,
			\end{equation}
			where $\dot{M}_{\text{CO}}$ is the CO formation rate, that is averaged over 2~Myr for each point and the factor 12/28 corrects for the mass of oxygen. 
			
Finally, the values of $f_{\text{CO}}$ for various resolutions are compared using the ratio
			\begin{equation}
				\label{eq:etaCO}
				\eta_{\text{CO}} = \frac{f_{\text{CO}}}{f_{\text{CO},\text{R}_\text{max}}} \, .
			\end{equation}
			
			\begin{figure*}
				\includegraphics[width=\linewidth]{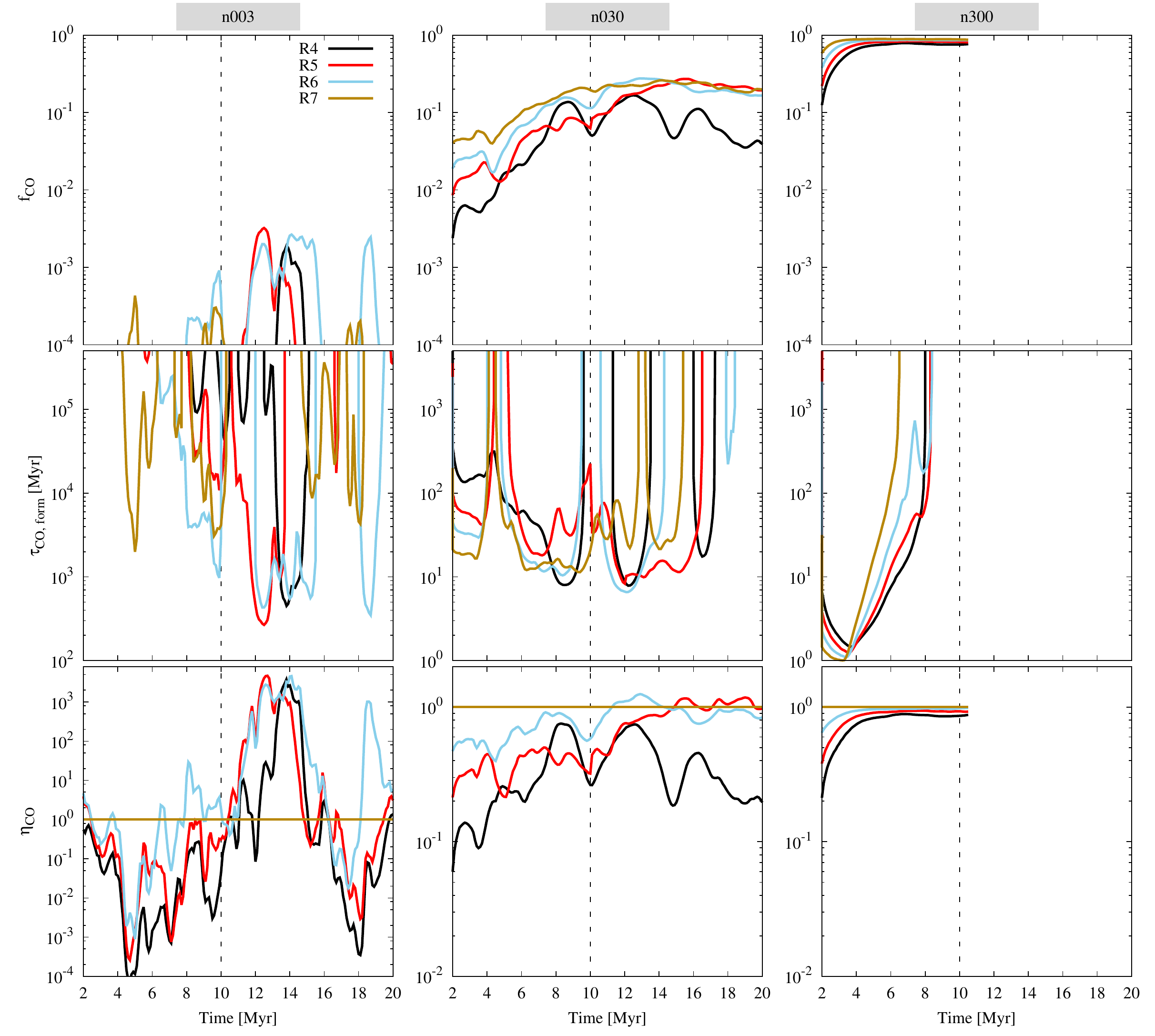}
				\caption{As in Figure~\ref{fig:TB_H2_analysis}, but now for CO. In the second row, $\tau_{\text{CO}, \text{form}} \rightarrow \infty$ when CO is being destroyed. Note the different extent of the y--axis in the second and third row. Similar to the results for H$_2$, the formation of CO is also not converged in all three density regimes.}
				\label{fig:TB_CO_analysis}
			\end{figure*}
			
Figure~\ref{fig:TB_CO_analysis} shows the evolution of $f_{\text{CO}}$ in the first row with a maximum value of 1 when 100\% of the C atoms are in CO. In the low density runs (first column), there is very little mass contained in the box, the mean density in the turbulent medium is low and any kind of shielding is ineffective, leading to insignificant CO abundances. Nevertheless, the differences with increasing resolution can still be investigated. There is no clear tendency of the changes in CO formation with regard to the resolution. For the intermediate density runs (second column), $f_{\text{CO}}$ increases and its fluctuations with time decrease as the resolution increases. The TB-R5-n030, TB-R6-n030, and TB-R7-n030 runs have similar $f_{\text{CO}}$ towards the end of the simulation, although their gas distribution is quite different, as shown in Figure~\ref{fig:TB_n30_coldens}. For the high density runs (third column), CO is initially rapidly produced. With increasing resolution, $f_{\text{CO}}$ saturates faster and at higher values, similar to H$_2$ in Figure~\ref{fig:TB_H2_analysis}. 
			
The second row shows the evolution of $\tau_{\text{CO}, \text{form}}$. Since CO is easily destroyed, multiple instances of $\tau_{\text{CO}, \text{form}} \rightarrow \infty$ is seen. In the low density runs, the evolution of $\tau_{\text{CO}, \text{form}}$ is chaotic since the environment suitable for CO formation is changing rapidly due to the turbulent stirring. In the intermediate density runs, the general trend of $\tau_{\text{CO}, \text{form}}$ with increasing resolution is difficult to discern but it lies between 10 and 1000~Myr for the various resolution runs. In the high density runs, $1 < \tau_{\text{CO}, \text{form}} < 10$~Myr during the early evolution at $t<4$~Myr; this  corresponds to the rapid initial increase in $f_{\text{CO}}$ seen in the first row. Notably, $\tau_{\text{CO}, \text{form}} < \tau_{\text{H}_2, \text{form}}$ during this phase, due to the artificial setup of the cold neutral medium, in which the equilibrium CO formation is found to be shorter than the equilibrium H$_2$ formation time-scale \citep[][and see Section~\ref{sec:CO_form_dissoc_time}]{Seifried2017b,Gong2018}. Similar to $\tau_{\text{H}_2, \text{form}}$, the value of $\tau_{\text{CO}, \text{form}}$ increases for higher resolution once $f_{\text{CO}}$ begins to saturate.
	
The third row shows the time evolution of $\eta_{\text{CO}}$ calculated with respect to R$_\text{max}=$R7. As expected, there are large, random variations in  $\eta_{\text{CO}}$ for the low density runs. In the intermediate density runs,  $\eta_{\text{CO}} \rightarrow 1$ with time for the TB-R5-n030 and TB-R6-n030 simulations, while the TB-R4-n030 run shows a large offset. The high density runs show that the large initial differences quickly settle down once $f_{\text{CO}}$ saturates and the $\eta_{\text{CO}} \approx 1$ at all resolutions towards the end of the runs. Again, similar to the H$_2$ formation, the evolution of total CO content is not converged for all resolutions ($\Delta x \geq 0.125$~pc ; see Section~\ref{sec:H2_CO_res_criteria} for explanation).
			
\subsubsection{Density distribution}
\begin{figure}
				\includegraphics[width=\linewidth]{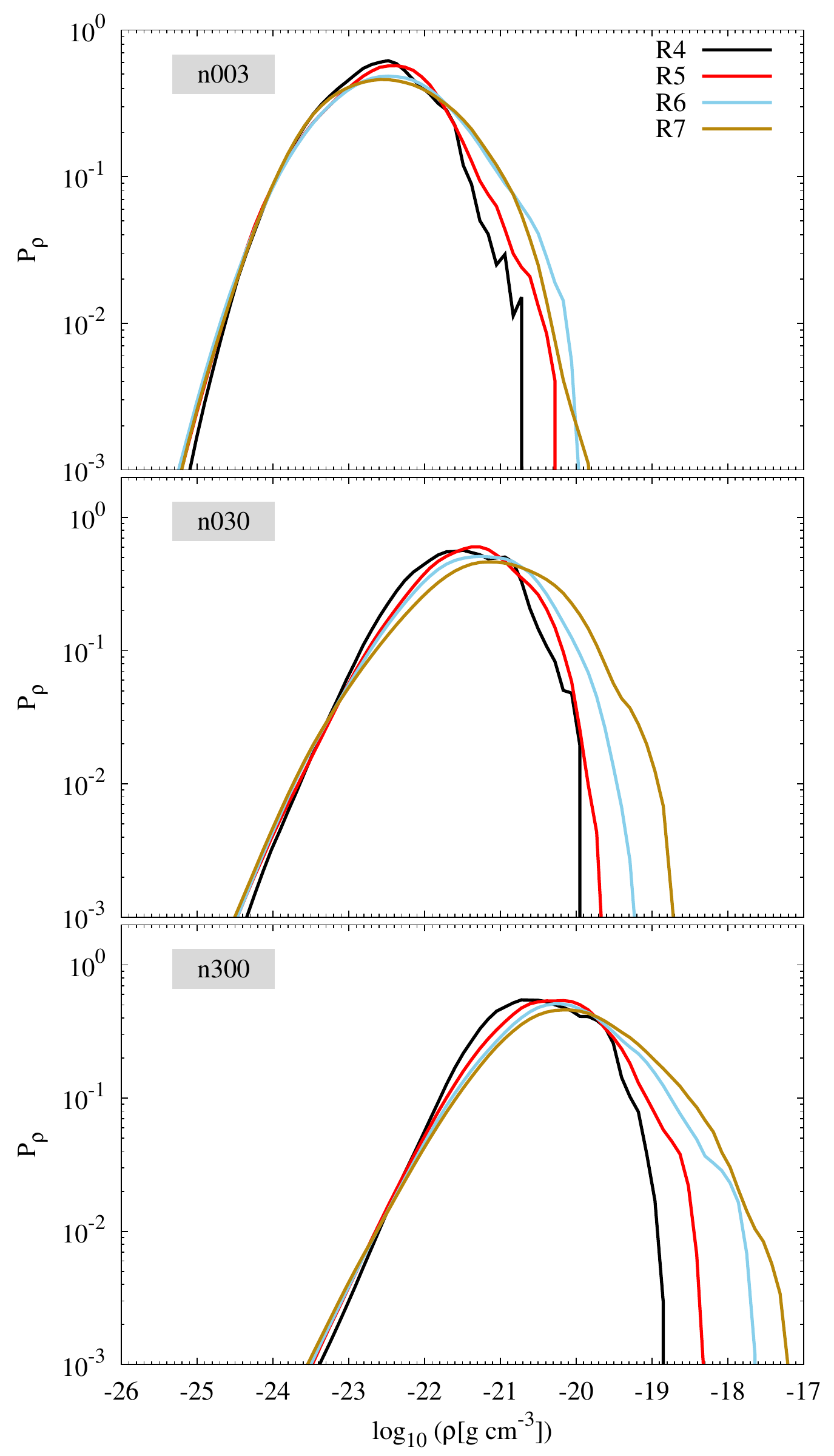}
				\caption{Mass-weighted PDFs (Equation~\ref{eq:dens_pdf}) of the gas density for the TB simulations with various resolutions. The distributions are averaged from five simulation snapshots between $t = 9.5$~Myr and 10~Myr. Low, intermediate and high density runs are shown from top to bottom. The density distributions are not converged in all three density regimes.}
				\label{fig:TB_1dpdf_dens}
			\end{figure}

The effective formation of CO molecules occurs in the dense regions, reasonably well-shielded from external radiation \citep{Glover10}. Thus, it is necessary to probe the mass-weighted density distribution at various resolutions, as shown in Figure~\ref{fig:TB_1dpdf_dens}. The mass contribution in the $i^\text{th}$ density bin is calculated as
			\begin{equation}
				M_i = \sum\limits_{j=1}^{\text{N}_{\text{cells},i}} \rho_j V_j \; ,
			\end{equation}
			where $\rho_j$ and $V_j$ are the density and volume of the $j^\text{th}$ cell, respectively, and $\text{N}_{\text{cells},i}$ are all cells in the computational domain with density $\rho$ such that 
			\begin{equation}
				\log \rho \in \left[ \log \rho_i , \log \rho_i + d(\log \rho_i) \right[ \; .
			\end{equation}
			where $d(\log \rho_i)$ is the bin size. Then, the mass-weighted density PDF, $P_\rho$, for every density bin $i$ is calculated as
			\begin{equation}
				\label{eq:dens_pdf}
				P_{\rho,i} = \frac{M_i}{M_\text{tot}} \, ,
			\end{equation}
			where $M_\text{tot}$ is the total mass in the simulation. 
			The distributions are averaged from five simulation snapshots between $t = 9.5$~Myr and 10~Myr, i.e. before gravity is activated. Hence, they display the density distributions in a medium with well-developed turbulence. The PDFs of low, intermediate, and high density runs are arranged from top to bottom. In all runs, the maximum density increases by a factor of 2 to 3 for each higher resolution. The low density tail is similar for all resolutions but the peak of the distribution as well as the high density tail of runs with different resolutions show significant differences. Thus, the mass-weighted distribution of the gas density is also not converged for the simulations with $\Delta x \geq 0.125$~pc (R7).
	

\subsection{Colliding Flow results} \label{sec:results_collflow}
\begin{figure}
\includegraphics[width=\linewidth]{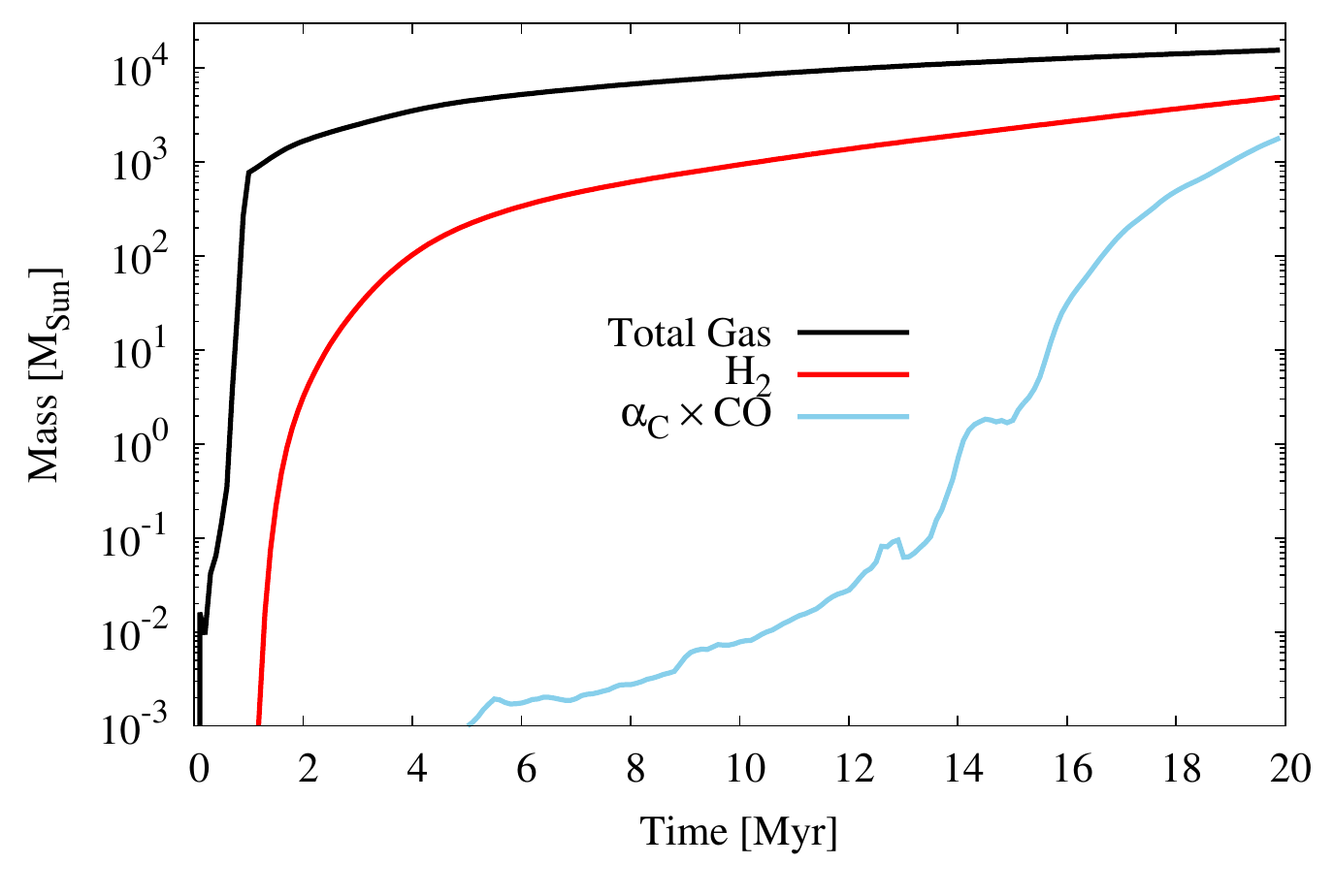} 
\caption{ Time evolution of the total, H$_2$, and CO mass in the CF-R10 run. Note that only the gas with  $T < 4000$~K is considered. The CO mass is scaled by $\alpha_\text{C}$ (see Equation~\ref{eq:carbon_factor}) in order to visually compare the formation of CO with H$_2$.} \label{fig:CF_mass_evol}
\end{figure}

Figure~\ref{fig:CF_mass_evol} shows the time evolution of the total gas mass with $T<4000$~K, H$_2$ and CO in the CF-R10 simulation with an effective resolution of $\Delta x = 0.008$~pc. The CO mass is scaled by $\alpha_\text{C}$ (see Equation~\ref{eq:carbon_factor}) to visually compare with the mass of the total gas and H$_2$. After about 1~Myr, the shocked gas at the collision interface starts to cool and rapid H$_2$ formation sets in. At the end of the simulation, $\sim 5 \times 10^3 \, \text{M}_\odot$ of H$_2$ has formed, which is about 30\% of the total mass. When the rapid H$_2$ formation slows down after $\sim$ 10\% of the gas mass is in H$_2$, the formation of CO starts. A significant amount of CO is produced only after 14~Myr. At the end of the simulation, about 3 $\text{M}_\odot$ of CO has formed,  which accounts for $\sim15\%$ of the total mass in carbon atoms.
		
\begin{figure*}
\includegraphics[width=0.98\linewidth]{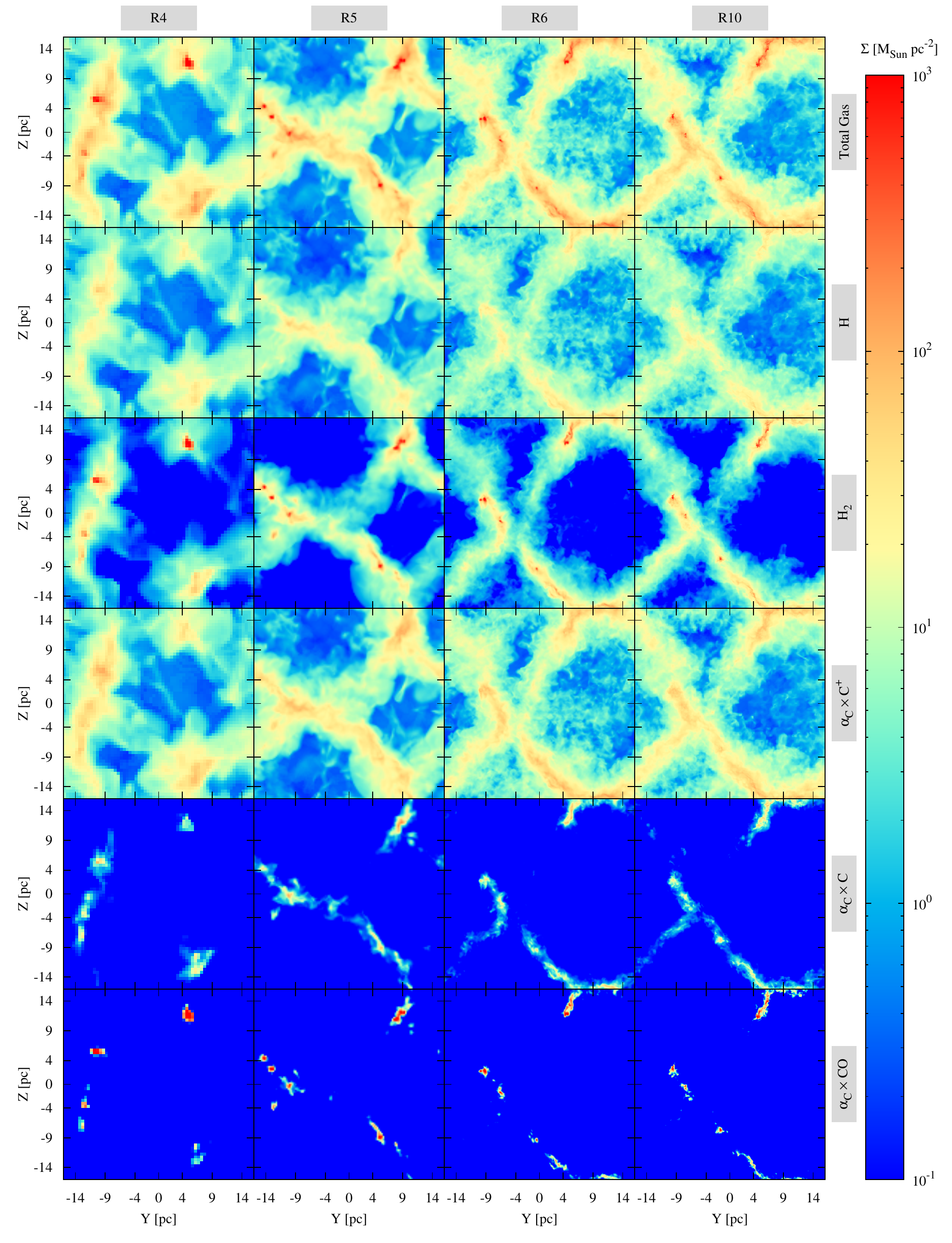} 
\caption{Column density distributions of the total gas, H, H$_2$, C$^+$, C and CO (top to bottom) at $t = 20$~Myr for the CF-R[4,5,6,10] runs (left to right). The C$^+$, C, and CO maps are scaled by $\alpha_\text{C}$ (see Equation~\ref{eq:carbon_factor}). The column density is computed by only considering the gas with $T < 4000$~K and integrating along the $x$--axis (i.e. the direction of the flow).} \label{fig:CF99_coldens}
\end{figure*}

Figure~\ref{fig:CF99_coldens} shows the column density of the total gas, H, H$_2$, C$^+$, C, and CO (top to bottom) at $t = 20$~Myr, integrated along the direction of the incoming flow ($x$--axis). Four selected resolutions (increasing from left to right) are displayed. The column densities of C$^+$, C and CO are scaled by $\alpha_\text{C}$ to make the carbon content visible on the same scale. The total gas column density shows that there is a significant change in the structure of the dense gas as the resolution increases in the CF-R[4-6] runs. However, the CF-R6 run and the highest resolution run (CF-R10) have similar gas distributions except for the densest regions. The clumpy and spatially separated dense regions obtained in the low resolution runs (CF-R4, CF-R5) convert to a network of filaments with compact cores at higher resolutions. The H column density shows the diffuse gas distribution that surrounds the denser H$_2$ regions. A diffuse C$^+$ distribution is seen throughout the simulation domain. The low and extended column density of atomic carbon surrounds the CO-rich regions. 
			
\subsubsection{Convergence study of H$_2$} \label{sec:CF_H2_conv_study}
\begin{figure*}
\includegraphics[width=\linewidth]{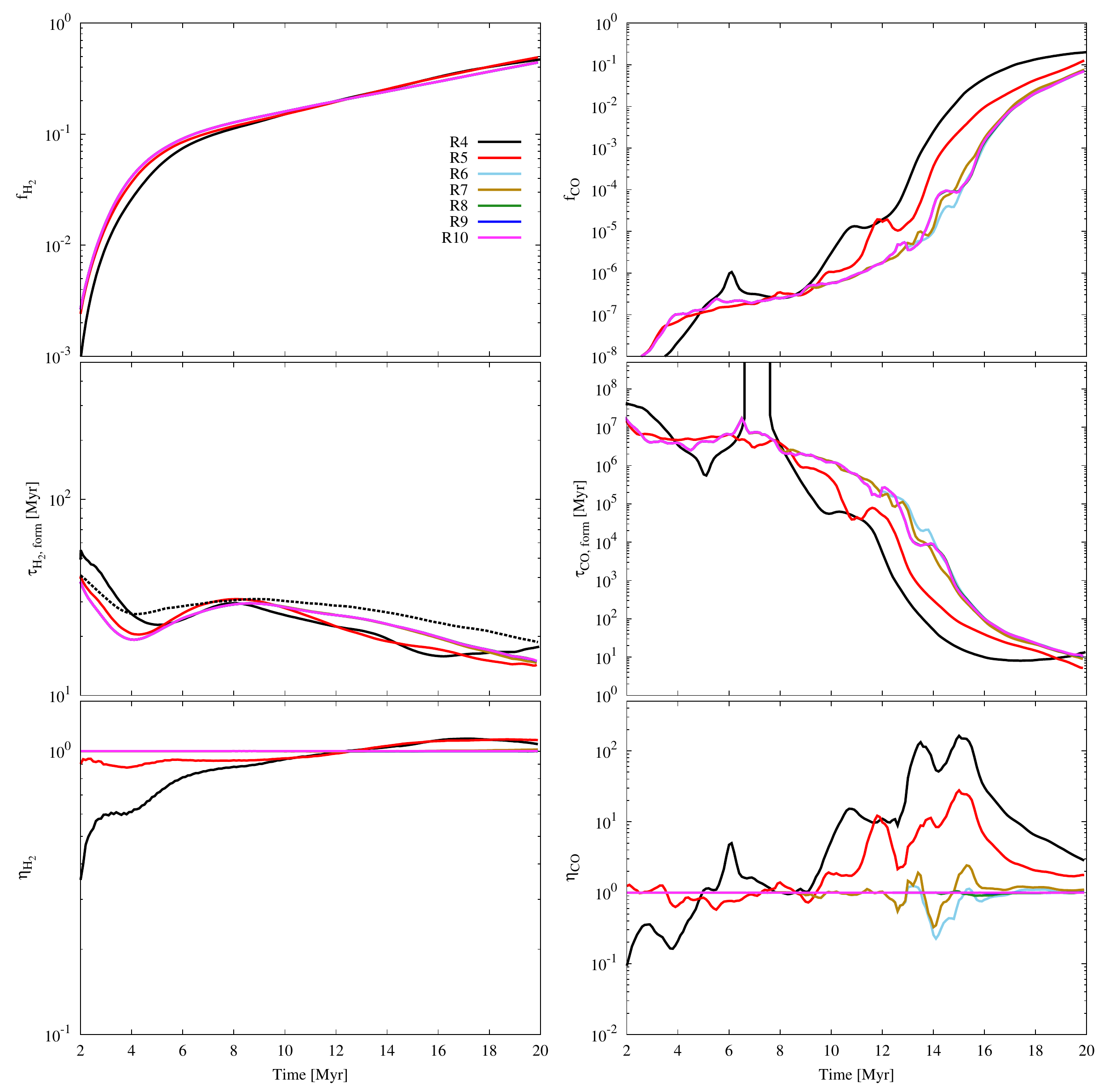} 
\caption{Time evolution of the molecular gas mass-fraction (top), formation time-scale (middle), and $\eta_{\text{H}_2}$ with respect to the highest resolution run CF-R10 for H$_2$ (left column) and CO (right column) in the CF setup. The dashed line in the middle-left panel shows the equilibrium H$_2$ formation time scale (see Eq.~\ref{eq:H2_formation}). The H$_2$ formation is converged for refinement level 6 with $\Delta x = 0.125$~pc (CF-R6) and higher. The CO formation is converged for refinement level 8 (CF-R8 simulation with $\Delta x = 0.032$~pc) and higher.}\label{fig:CF_H2_CO_analysis}
\end{figure*}

The left column of Figure~\ref{fig:CF_H2_CO_analysis} shows the evolution of H$_2$ in the CF runs. Similar to the results of the TB runs, the evolution of the molecular content is shown starting from 2~Myr. The top panel shows the evolution of $f_{\text{H}_2}$ (Equation~\ref{eq:fH2}). About 40\% of H atoms are found in H$_2$ towards the end of the simulations. Compared to the higher resolution runs, the CF-R4 and CF-R5 runs have lower $f_{\text{H}_2}$ before 10~Myr and then higher $f_{\text{H}_2}$ towards the end of the simulation. The values of $f_{\text{H}_2}$ in CF-R6 and higher resolution runs are similar.
        
The middle panel shows the evolution of  $\tau_{\text{H}_2, \text{form}}$ (Equation~\ref{eq:tauH2form}). The dashed line depicts the equilibrium H$_2$ formation time scale $\langle \tau_{\text{H}_2,\text{eq}} \rangle_\text{mass}$ given by Equation~\ref{eq:H2_formation}. After the rapid H$_2$ formation phase (i.e. $t<6$~Myr), $\tau_{\text{H}_2, \text{form}}$ is always shorter than $\langle \tau_{\text{H}_2,\text{eq}} \rangle_\text{mass}$. As the mean density of the MC forming gas gradually increases with time, $\tau_{\text{H}_2,\text{form}}$ decreases to $< 20$~Myr towards the end of the simulation. The CF-R4 and CF-R5 runs have longer $\tau_{\text{H}_2,\text{form}}$ than the other runs during the early evolutionary phase, but have shorter $\tau_{\text{H}_2,\text{form}}$ after $t \sim 10$~Myr. There are only minor differences in the evolution of $\tau_{\text{H}_2,\text{form}}$ for the other high resolution runs.
        
The bottom panel shows the time evolution of $\eta_{\text{H}_2}$ (Equation~\ref{eq:etaH2}) in order to compare  $f_{\text{H}_2}$ in the runs with various resolutions. Here, the reference simulation is CF-R10 with R$_\text{max}=$~R10, corresponding to an effective resolution of $\Delta x = 0.008$~pc. The CF-R4 and CF-R5 runs have $\eta_{\text{H}_2}<1$ at early times but eventually exceed the amount of H$_2$ produced by the higher resolution runs ($\eta_{\text{H}_2}>1$) towards the end of the simulation. There is no significant difference in $f_{\text{H}_2}$ between the other higher resolution runs.

Overall, the H$_2$ formation is not sensitive to the (Jeans-)refinement of the dense regions; the total H$_2$ abundance and average H$_2$ formation time-scale are converged for an effective resolution of $\Delta x = 0.125$~pc (CF-R6) and higher. This is in agreement with the required resolution for H$_2$ chemistry as noted by \citet{Glover10,Micic12, Valdivia2016} and \citet{Seifried2017b} (see Section~\ref{sec:H2_CO_res_criteria}).

\subsubsection{Convergence study of CO} \label{sec:CF_CO_conv_study}
The right column of Figure~\ref{fig:CF_H2_CO_analysis} shows the evolution of CO in the CF runs. The top panel shows the evolution of $f_\text{CO}$ (Equation~\ref{eq:fCO}). The CF-R4 run generally has lower $f_\text{CO}$ than the other higher resolution runs in the beginning. With time, $f_\text{CO}$ becomes significantly larger for the low resolution CF-R4 and CF-R5 runs. The spikes seen in the low resolution runs (e.g. in CF-R4 at $t = 6$~Myr) are absent in higher resolution runs. Between $ t = 12$ and $16$~Myr, the CF-R6 and CF-R7 runs show some deviation in $f_\text{CO}$ from the higher resolution runs.

The middle panel shows the evolution of $\tau_{\text{CO, form}}$ (Equation~\ref{eq:tauCOform}). Only the CF-R4 run shows $\tau_{\text{CO, form}} \rightarrow \infty$ when a significant amount of CO is destroyed after $t\sim6$~Myr. Before 10~Myr, the CO formation time-scale is extremely long. Afterwards, the CF-R4 and CF-R5 runs have shorter $\tau_{\text{CO,form}}$ than the higher resolution runs. With time, the differences in $\tau_{\text{CO,form}}$ at various resolutions become smaller and all the runs have $\tau_{\text{CO,form}} \sim$ 10~Myr towards the end of the simulations.

The bottom panel shows the time evolution of $\eta_{\text{CO}}$ (Equation~\ref{eq:etaCO}, with R$_\text{max}=$R10). Before 5~Myr, the CF-R4 and CF-R5 runs have $\eta_{\text{CO}}<1$, but they quickly produce as much CO as the CF-R10 run. After 10~Myr, $\eta_{\text{CO}}\gg1$ for the CF-R4 and CF-R5 runs. With increasing resolution, the deviation from the reference value decreases. The runs CF-R6 and CF-R7 show fluctuations with $0.2<\eta_{\text{CO}}<2$ and only the CF-R8 and CF-R9 runs show $\eta_{\text{CO}}\approx1$. Hence, the sub-structure of small, dense regions significantly affects the CO formation; the total CO abundance and the average CO formation time-scale are converged for an effective resolution of $\Delta x = 0.032$~pc (CF-R8) and higher.

\begin{figure}
\includegraphics[width=\columnwidth]{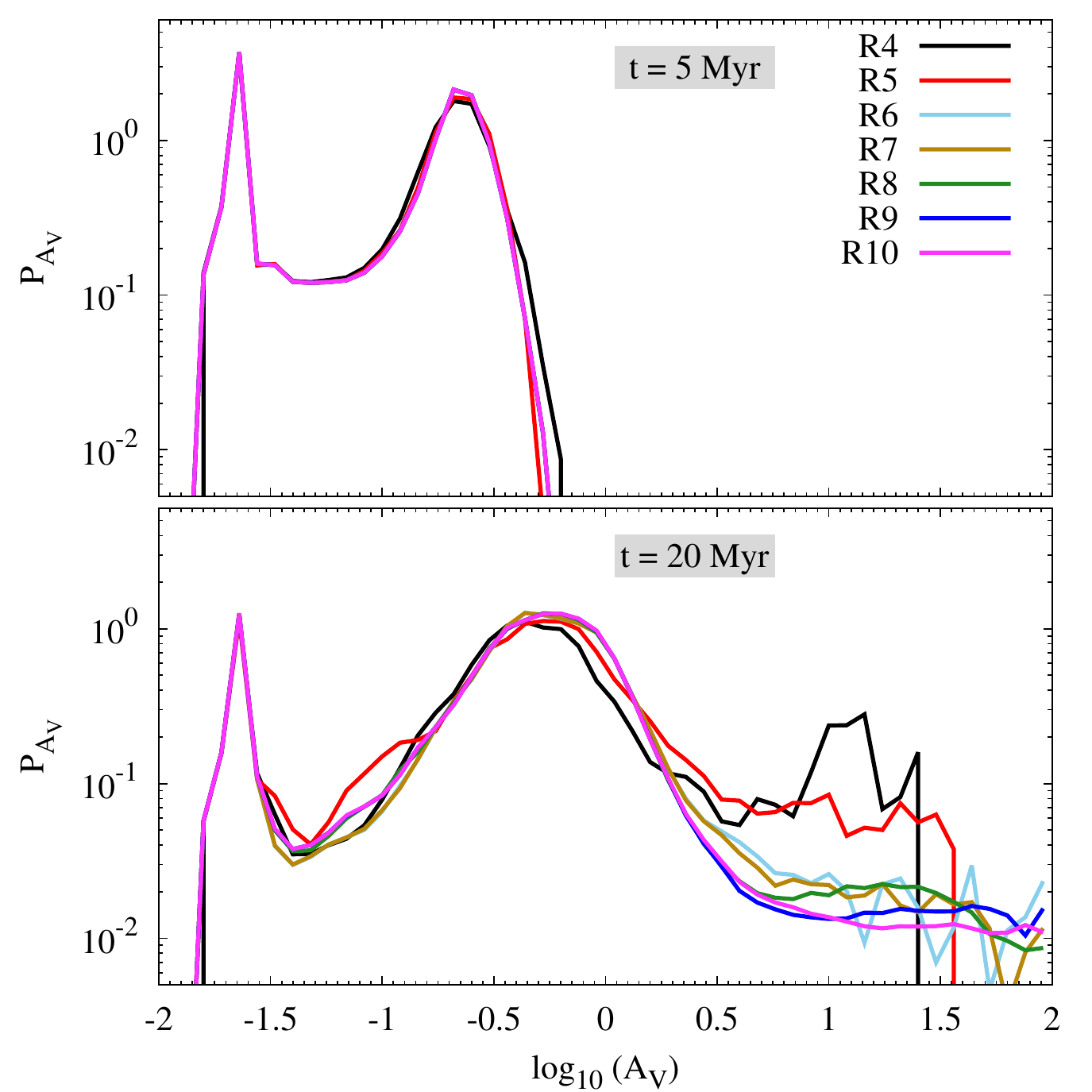}
\caption{The mass-weighted $A_V$-PDFs of the CF runs with different resolutions. Top: at time $t=5$~Myr, bottom: at  $t=20$~Myr. Lower resolution runs eventually produce more gas with $A_V>1$ and hence more molecular gas.}\label{fig:CF_pdf_evol}
\end{figure}

The formation of a higher amount of H$_2$ and CO in lower resolution simulations is explained by the distribution of the visual extinction, $A_V$, experienced by the cells in the simulation, shown in Figure~\ref{fig:CF_pdf_evol}. For each cell, $A_V$ is the average value over the 48 \textsc{Healpix} pixels, computed by the \textsc{OpticalDepth} module (see Section~\ref{sec:opticaldepth}). The top panel shows the $A_V$-PDF at $t=5$~Myr and the bottom panel at $t=20$~Myr, when both H$_2$ and CO are being produced rapidly. It is clearly seen that the mass of well-shielded gas (around $A_V=1$) is similar at early times; however, at $t=20$~Myr, the shielding in lower resolution runs (CF-R4 and CF-R5) has become more effective than in the higher resolution runs. Indeed, the column densities of the CF runs in Figure~\ref{fig:CF99_coldens} show that a coarser resolution produces large, clumpy structures that eventually lead to higher shielding of the gas within, resulting in a more effective molecule formation. As more molecules form, the non-linear effects due to (self--)shielding and density further increase the molecular content of the cold gas.
	
\subsubsection{Density distribution} \label{sec:CF_dens_pdf}
\begin{figure}
\includegraphics[width=\linewidth]{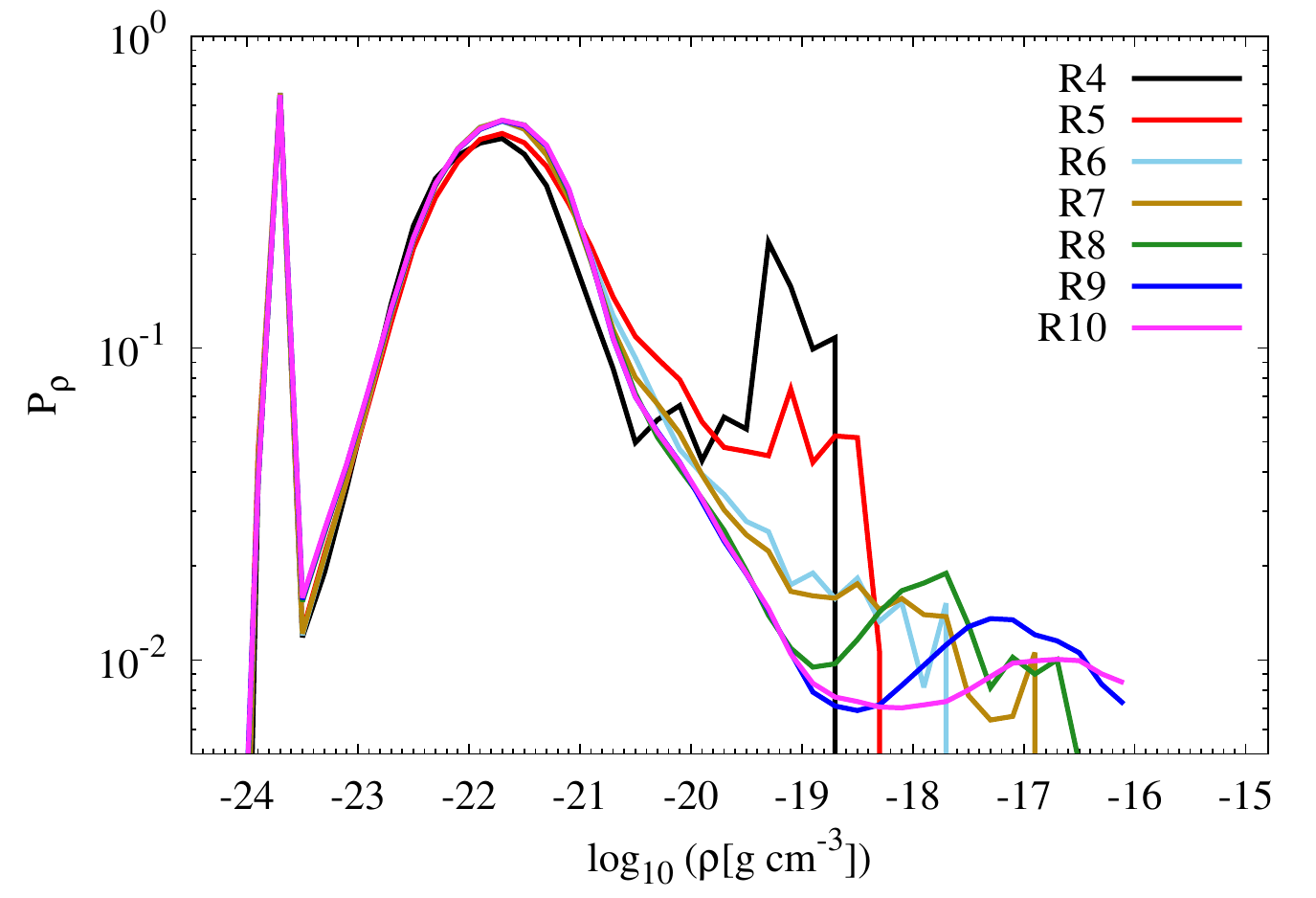}
\caption{Mass-weighted density PDFs (Equation~\ref{eq:dens_pdf}) of the CF runs with various resolutions. The distributions are averaged from 5 simulation snapshots between $t=19.5$~Myr and 20~Myr. The high density tails show significant differences with increasing resolution and the density at which the distributions deviate from one another increases with increasing resolution.}
		\label{fig:CF_1dpdf_dens}
	\end{figure}

Figure~\ref{fig:CF_1dpdf_dens} shows the mass-weighted density PDFs (Equation~\ref{eq:dens_pdf}) of the CF simulations with various resolutions. The distributions are averaged from five simulation snapshots between $t=19.5$~Myr and 20~Myr. The spike at low density ($\rho \sim 10^{-24}$~g~\pcmc) belongs to the incoming warm gas that begins to cool below 4000~K near the collision interface. The distribution of the gas below the density of $\rho \sim 10^{-22}$~g~\pcmc (peak of the distribution) is similar for all the resolutions. The high density tails in Figure~\ref{fig:CF_1dpdf_dens} clearly show the effect of resolution on the distribution of gas. With increasing resolution, the density structures of interest  -- i.e. shocks or gravitating regions -- are better resolved and higher densities are obtained. This occurs because the gas is collected in relatively large cells in a low resolution run, whereas the same gas is compressed into smaller sub-structures for higher resolution runs. It is also apparent that the density at which the PDFs start to deviate from one another increases with increasing resolution. This influences the convergence of the molecular gas mass in the presented runs and will be discussed in the next section.
	

\section{The resolution criterion for the convergence of H$_2$ and CO formation}
	\label{sec:H2_CO_res_criteria}
	Qualitatively, the required spatial resolution for converged H$_2$ formation in multiple chemical networks has been shown to be $\Delta x_{\text{H}_2,\text{form}} \sim 0.1$~pc by \citet{Glover10}, \citet{Micic12}, \citet{Valdivia2016}, and \citet{Seifried2017b}; the CF runs agree with this requirement. A theoretical foundation, however, is still missing. Here, the resolution requirements for H$_2$ and CO formation are discussed using the relevant time-scales and densities.

Three important time-scales should be considered to understand the chemical evolution of the gas, namely the molecule formation time-scale ($\tau_{\text{mol},\text{form}}$), the molecule dissociation time-scale ($\tau_{\text{mol},\text{dissoc}}$), and the local cell crossing time-scale of the gas ($t_\text{cell,cross}$). The relations between these time-scales result in two conditions relevant for the converged molecule formation in a simulation. 

Condition 1 is a \textbf{physical condition} and is given by
\begin{equation}\label{eq:physical_condt}
\tau_{\text{mol},\text{form}} \leq \tau_{\text{mol},\text{dissoc}},
\end{equation}
such that effective molecule formation is possible. If a simulation is not able to resolve the typical density at which Condition 1 is satisfied, then the qualitative correctness of the molecule formation is questionable; i.e. regions that should contain molecular gas are not likely to have any molecules (see Section~\ref{sec:form_dissoc_time}).

Condition 2 is a \textbf{dynamical condition} and is given by
\begin{equation}\label{eq:dynamical_condt}
\tau_{\text{mol},\text{form}} \leq t_\text{cell,cross} \, .
\end{equation}

	If a simulation resolves the typical densities that are high enough to satisfy Condition 1 but too low to satisfy Condition 2, then the molecule formation may be qualitatively correct, but quantitatively incorrect; i.e. the simulation will correctly predict that the gas in a certain region should be molecular, but might get the molecular fraction or the time to reach the equilibrium wrong. Cells that fulfill the dynamical condition become fully molecular within one cell crossing time regardless of the physical properties of the gas. The dynamical condition is important because it addresses the fact that small dynamical changes in the environment easily change the history of molecule formation in a cell due to the non-linear nature of the turbulent gas evolution (see Section~\ref{sec:crossing_time}).  

Both conditions are applied to our simulation results in Sections~\ref{sec:CF_res_req}~and~\ref{sec:TB_res_req}.

\subsection{Condition 1: Physical condition} \label{sec:form_dissoc_time}
The physical condition (Eq.~\ref{eq:physical_condt}) equates the molecule formation and dissociation time scale, which can be investigated for any given molecule. It results in a density criterion, indicating which densities need to be resolved by simulation. Here we calculate the density criterion only for H$_2$ and CO. For both molecules, we show that the resolution requirement following from Eq.~\ref{eq:physical_condt} is weaker than the one following from Eq.~\ref{eq:dynamical_condt}, but this could be different for more complex molecules.

\subsubsection{H$_2$ molecule} \label{sec:H2_form_dissoc_time}		
The H$_2$ formation occurs predominantly on the surface of dust grains \citep{Hollenbach1989} via the reaction
\begin{equation}
\text{H} + \text{H(s)} \longrightarrow \text{H}_2 \, ,
\end{equation}
with the rate coefficient $k_{165}$ in \citet{Glover10}. Photo-dissociation is the main destruction mechanism of H$_2$ molecules \citep[e.g][]{Draine1996} via the reaction
\begin{equation}
\text{H}_2 + \gamma \longrightarrow \text{H} + \text{H} \, ,
\end{equation}
with the H$_2$ photodissociation rate given by
\begin{equation}\label{eq:H2_dissoc_rate}
R_{\text{pd,H}_2} = 3.3 \times 10^{-11} \, G_0 \, f_{\text{dust,H}_2} \, f_{\text{shield,H}_2} \text{ s}^{-1} \, ,
\end{equation}
where $G_0=1.7$ in Habing units denotes the strength of the interstellar radiation field. The factors $f_{\text{dust,H}_2}$ and $f_{\text{shield,H}_2}$ account for the dust extinction and H$_2$ self-shielding, respectively. If $f_{\text{dust,H}_2} = 0$, the radiation incident on the H$_2$ molecules is completely attenuated by dust. Similar holds for $f_{\text{shield,H}_2}$ which is a function of H$_2$ column density and gas temperature \citep[see][for further details]{Glover10,Walch15_SILCC1}. Using these reaction rates, the distribution of the H$_2$ formation time-scale ($\tau_{\text{H}_2,\text{form}}$) and H$_2$ dissociation time-scale ($\tau_{\text{H}_2,\text{dissoc}}$) in a simulation can be obtained.

\begin{figure}
\includegraphics[width=\columnwidth]{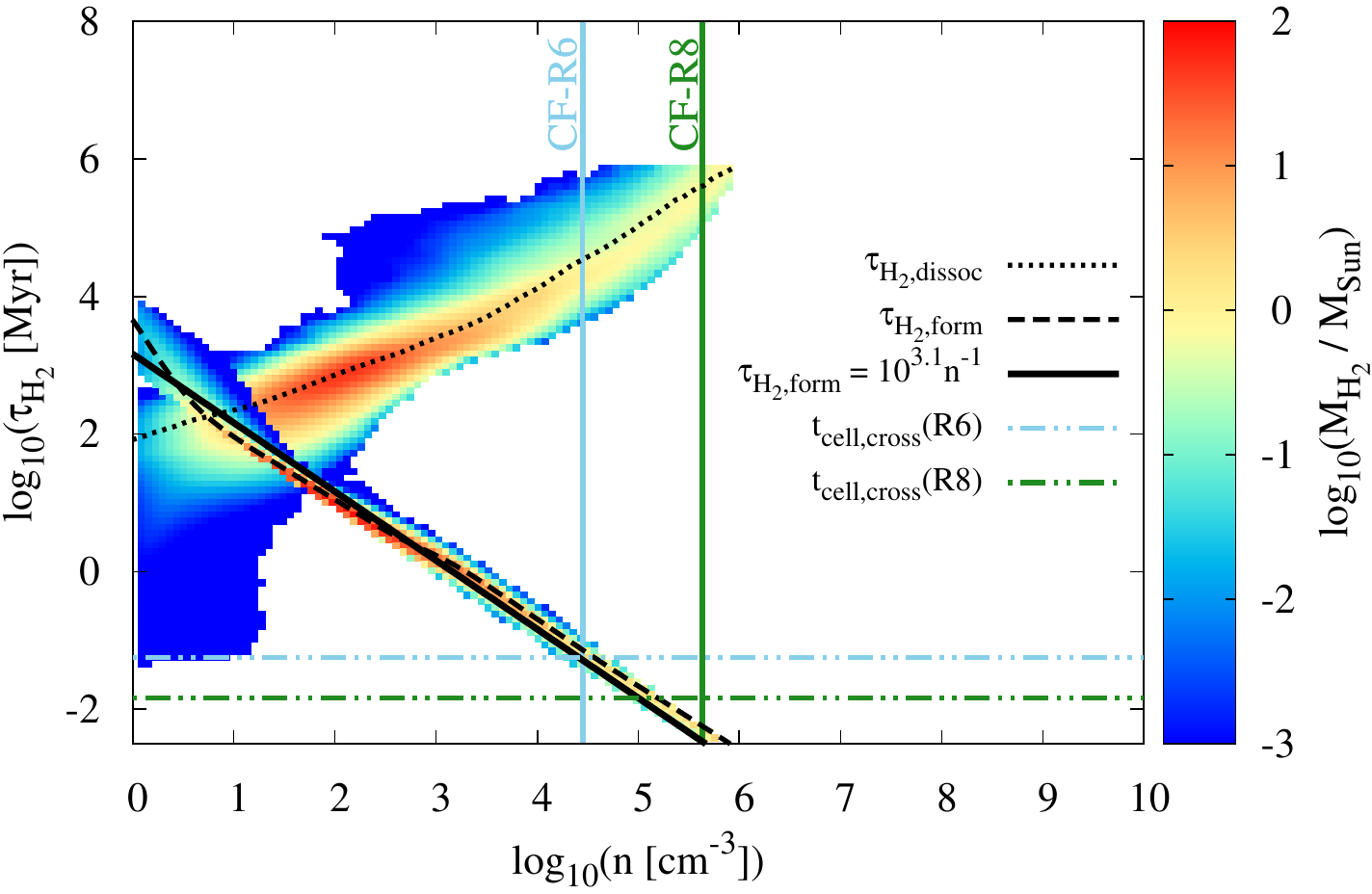}
\caption{The distribution of the H$_2$ formation time-scale ($\tau_{\text{H}_2,\text{form}}$) and dissociation time-scale ($\tau_{\text{H}_2,\text{dissoc}}$) as a function of the number density of the cells in the highest resolution CF-R10 run at $t=20$~Myr. The colour code denotes the mass of H$_2$. The dashed and dotted black lines denote the H$_2$-mass-weighted mean of the distributions. The solid, black line represents the power law fit for $\tau_{\text{H}_2,\text{form}}(n)$. The coloured, horizontal, dot-dashed lines show $t_\text{cell,cross}$ (Equation~\ref{eq:t_cell_cross}, with $\langle \sigma \rangle_\text{mass}=2$~\kms) for two different resolution runs -- cyan: CF-R6 and green: CF-R8. The coloured, vertical, solid lines refer to the Jeans density corresponding to the spatial resolution of the CF-R6 and CF-R8 runs, as given by Equation~\ref{eq:n_Jeans} for cold gas.} \label{fig:2dpdf_time_dens_H2}
\end{figure}

The 2D-histogram in Figure~\ref{fig:2dpdf_time_dens_H2} shows how the two time-scales, $\tau_{\text{H}_2,\text{form}}$ and $\tau_{\text{H}_2,\text{dissoc}}$ are related with the gas number density in the CF-R10 run at $t=20$~Myr. The colour code shows the H$_2$ mass distribution. The H$_2$-mass-weighted mean of the each distribution is displayed by the black, dashed ($\tau_{\text{H}_2,\text{form}}$) and dotted ($\tau_{\text{H}_2,\text{dissoc}}$) lines. Cells with densities $n>10$~\pcmc have $\tau_{\text{H}_2,\text{form}}< \tau_{\text{H}_2,\text{dissoc}}$ and thus, form H$_2$ faster than it is dissociated\footnote{Since the photodissociation rate depends linearly on the incident radiation field, the required density will increase in environments with higher $G_0$, for instance near star-forming regions.}. As a result, almost all of the H$_2$ is found in cells that fulfill the physical condition
	\begin{equation} \label{eq:nH2_physical}
			n_{\text{H}_2,\text{physical}}\gtrsim10\text{~\pcmc}\, .
	\end{equation}

The solid, black line in Figure~\ref{fig:2dpdf_time_dens_H2} denotes the linear fit for $\log_{10}\tau_{\text{H}_2,\text{form}}$($\log_{10} n$) for the cells that satisfy the physical condition. This leads to the power law relation
		\begin{equation}
			\label{eq:H2_form_fit}
			\tau_{\text{H}_2,\text{form}} = 10^{3.1} \text{ Myr} \left( \frac{n}{\text{1~\pcmc}} \right)^{-1} \, ,
		\end{equation}		 
	which is similar to the equilibrium H$_2$ formation time-scale ($\tau_{\text{H}_2,\text{eq}}$,  Equation~\ref{eq:H2_formation}).
	
\subsubsection{CO molecule}
	\label{sec:CO_form_dissoc_time}
		\begin{figure}
			\includegraphics[width=\columnwidth]{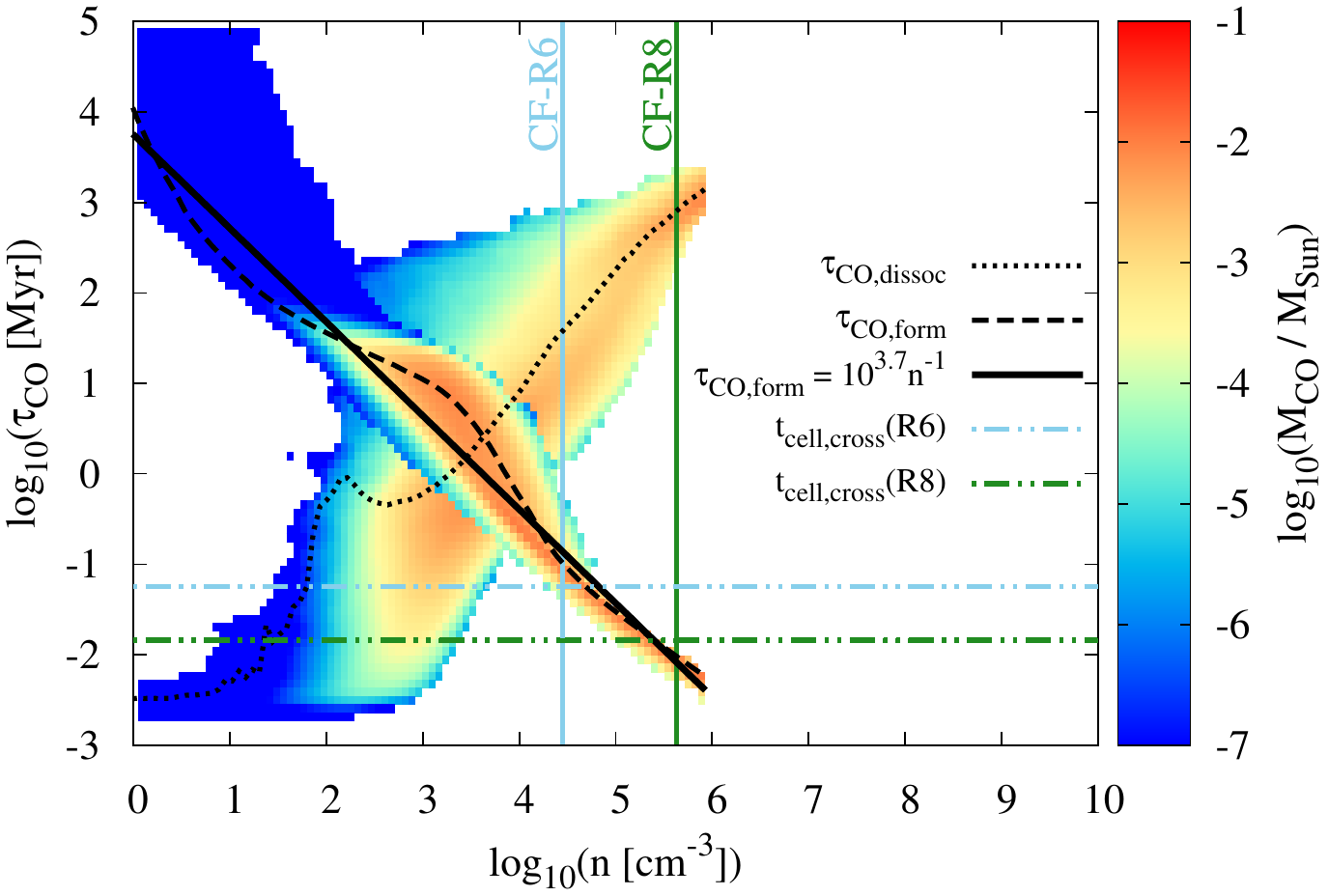}
			\caption{As in Figure~\ref{fig:2dpdf_time_dens_H2} but now for CO. The solid, black line represents the power law fit for $\tau_{\text{CO},\text{form}}(n)$ for densities $n > 10^4$~\pcmc that fulfill the physical condition (Equation~\ref{eq:physical_condt}). The dynamical condition (Equation~\ref{eq:dynamical_condt}) is fulfilled at a much higher density associated with the CF-R8 run.} 
			\label{fig:2dpdf_time_dens_CO}
		\end{figure}
		The shielding of CO by H$_2$ is important for CO formation. Therefore, effective CO formation sets in only when enough H$_2$ is formed \citep[][see also Figure~\ref{fig:CF_mass_evol}]{Glover10} and hence the formation of H$_2$ on dust grains is a limiting reaction for CO formation. In addition, the formation of CH$_2^+$ via the reactions
		\begin{flalign}
		\label{eq:CH2p_reaction1}
			&\text{C}^+ + \text{H}_2 \longrightarrow \text{CH}_2^+ + \gamma \, ,\\
			\label{eq:CH2p_reaction2}
			&\text{C} + \text{H}_3^+ \longrightarrow \text{CH}_2^+ + \text{H} \, ,\\
			&\text{H}_2 + \text{cosmic-ray} \longrightarrow \text{H}_2^+ + \text{e}^- \, ,\\
			\label{eq:H3p_dissoc}
			&\text{H}_3^+ + \text{e}^- \longrightarrow \text{H}_2 + \text{H} \, ,
 		\end{flalign}
 		have been identified as the limiting reactions for CO formation \citep[see][for further details]{Glover2012chemcompare}. The net formation rate of CH$_2^+$ is then given by \citet[][]{Glover2012chemcompare} as
 		\begin{equation}
 			R_{\text{CH}_2^+} = k_{\text{CH}_2^+,1} n_{\text{C}^+} \, n_{\text{H}_2} + 2\, \zeta_{\text{H}} \, n_{\text{H}_2} \frac{k_{\text{CH}_2^+,2}}{k_{\text{dr}}} \frac{n_\text{C}}{n_{\text{e}^-}} \,
 		\end{equation}
 		where $k_{\text{CH}_2^+,1}$ denotes the rate coefficient for reaction~\ref{eq:CH2p_reaction1}, $n_x$ denotes the number density of the chemical species $x$, $\zeta_{\text{H}}=3\times10^{-17} \text{ s}^{-1}$ is the cosmic ray ionisation rate of atomic hydrogen, $k_{\text{CH}_2^+,2}$ denotes the rate coefficient for reaction~\ref{eq:CH2p_reaction2}, and $k_{\text{dr}}$ denotes the rate coefficient for the dissociative recombination reaction~\ref{eq:H3p_dissoc}.
		
When the cosmic ray \citep{Bisbas2015} and X-ray \citep{Mackey2018} ionisation rates are moderate, the CO molecules are mostly destroyed by photodissociation via the reaction
		\begin{equation}
			\text{CO} + \gamma \longrightarrow \text{C} + \text{O} \, ,
		\end{equation}
		with the reaction rate
		\begin{equation}
			\label{eq:CO_dissoc_rate}
			R_{\text{pd,CO}} = 1.235 \times 10^{-10} \, G_0 \, f_\text{dust,CO} \, f_\text{shield,CO} \text{ s}^{-1} \, ,
		\end{equation}
		where the factor $f_\text{dust,CO}$ accounts for the shielding of CO by the dust and $f_\text{shield,CO}$  incorporates the shielding of the CO by H$_2$ as well as CO self-shielding \citep[see][for further details]{Glover10,Walch15_SILCC1}. Using these reaction rates, the CO formation time-scale ($\tau_{\text{CO},\text{form}}$) and dissociation time-scale ($\tau_{\text{CO},\text{dissoc}}$) can now be determined.

As for H$_2$, the 2D-histograms in Figure~\ref{fig:2dpdf_time_dens_CO} show the distribution of the two time-scales, $\tau_{\text{CO},\text{form}}$ and $\tau_{\text{CO},\text{dissoc}}$, as a function of the gas number density in the CF-R10 run at $t=20$~Myr. The colour code denotes the mass of CO in each bin. The CO-mass-weighted mean of the each distribution is displayed by the black, dashed ($\tau_{\text{CO},\text{form}}$) and dotted ($\tau_{\text{CO},\text{dissoc}}$) lines. Given the significant amount of CO in cells with  $n\gtrsim10^3$~\pcmc, it can be concluded that the physical condition for CO chemistry is satisfied by cells with 
	\begin{equation} \label{eq:nCO_physical}
			n_{\text{CO},\text{physical}}\gtrsim10^3\text{~\pcmc}\, .
	\end{equation}

The solid, black line in Figure~\ref{fig:2dpdf_time_dens_CO} shows the linear fit of $\log_{10}\tau_{\text{CO},\text{form}}$($\log_{10} n$) for the cells that satisfy the physical condition. This leads to the power law relation
			\begin{equation}
			\label{eq:CO_form_fit}
				\tau_{\text{CO},\text{form}} = 10^{3.7} \text{ Myr} \, \left( \frac{n}{1 \text{ cm}^{-3}} \right)^{-1} \, .
			\end{equation}
Thus, the typical CO formation time-scale is slightly longer than the H$_2$ formation time-scale (Equation~\ref{eq:H2_form_fit}).

\subsection{Condition 2: Dynamical condition}\label{sec:crossing_time}
It is evident that the physical condition in Equation~\ref{eq:physical_condt} is a necessary but not a sufficient condition for converged molecule formation. With the expression for the typical H$_2$ and CO formation time at hand (Equations~\ref{eq:H2_form_fit}~and~\ref{eq:CO_form_fit}), the dynamical condition, which equates the molecule formation and cell crossing time-scales (see Equation~\ref{eq:dynamical_condt}), can be examined once the typical cell crossing time of the dense gas is determined.
	
The local (M)HD time step is given by \citep{Courant1967}
	\begin{equation}
	\label{eq:t_CFL}
		\Delta t = \text{CFL}\times \frac{\Delta x}{v}\, ,
	\end{equation}
	where $\Delta x$ is the size of the grid cell, $v$ is the velocity of the gas in the cell, and the constant factor CFL $\leq 1$ \citep[see e.g.][]{Derigs2016}. In the following discussion, CFL$ =1$ is taken for simplicity. Furthermore, determining $\Delta t$ via the velocity of individual cells is a conservative estimate. The reason is that the chemical abundances are advected with the flow and therefore large advection velocities do not change the density distribution of the gas. Since molecule formation occurs in dense and turbulent gas, it is intuitive to take the mass-weighted mean velocity dispersion $\langle \sigma \rangle_\text{mass}$ of such dense gas as an estimate for the typical velocity in molecule forming gas. This results in the typical cell crossing time of the gas, defined as
	\begin{equation}
	\label{eq:t_cell_cross}
		t_\text{cell,cross} = \frac{\Delta x}{\langle \sigma \rangle_\text{mass}}\, .
	\end{equation}
	
Substituting the time-scales from Equations~\ref{eq:H2_form_fit}~and~\ref{eq:t_cell_cross} in Equation~\ref{eq:dynamical_condt}, the gas number density which fulfills condition 2 for the H$_2$ molecule is
		\begin{equation}
			\label{eq:nH2_dynamical}
			n_{\text{H}_2,\text{dynamical}} \geq 10^{3.1}\text{ cm}^{-3} \left(\frac{\langle \sigma \rangle_\text{mass}}{\text{1 km s}^{-1}}\right)  \left( \frac{\Delta x}{\text{1 pc}}\right)^{-1} \, .
		\end{equation}
Similarly, from Equations~\ref{eq:CO_form_fit}, \ref{eq:t_cell_cross}~and~\ref{eq:dynamical_condt}, the gas number density which fulfills condition 2 for the CO molecule is found to be
		\begin{equation}
			\label{eq:nCO_dynamical}
			n_{\text{CO},\text{dynamical}} \geq 10^{3.7}\text{ cm}^{-3} \left(\frac{\langle \sigma \rangle_\text{mass}}{\text{1 km s}^{-1}}\right)  \left( \frac{\Delta x}{\text{1 pc}}\right)^{-1} \, .
		\end{equation}
Therefore, the density required for the CO formation process to fulfill the dynamical condition is $\sim4$ times higher than that for H$_2$. The densities in Equations~\ref{eq:nH2_dynamical}~and~\ref{eq:nCO_dynamical} are denoted by the intersection of the solid, black line and horizontal,dot-dashed lines in Figures~\ref{fig:2dpdf_time_dens_H2}~and~\ref{fig:2dpdf_time_dens_CO}, respectively.

\subsection{Resolution requirements for the CF simulations} \label{sec:CF_res_req}
The density estimates such as in Equations~\ref{eq:nH2_dynamical}~and~\ref{eq:nCO_dynamical} are sufficient for the theoretical understanding of the resolution criteria. However, it is practical to have an estimate of the spatial resolution required to simulate the formation of a molecular cloud with converged H$_2$ and CO formation. For simulations with self-gravitating gas (e.g. the CF runs), the spatial resolution associated with each density in Equations~\ref{eq:nH2_dynamical}~and~\ref{eq:nCO_dynamical} can be calculated via the Jeans density \citep{Jeans1902}
		\begin{equation}
			\label{eq:rho_Jeans}
			\rho_\text{Jeans} = \frac{\pi c_\text{s} ^2}{\lambda_\text{J}^2 G} \, ,
		\end{equation}

		where $\lambda_\text{J}=N_\text{J} \, \Delta x$ is the Jeans length resolved with $N_\text{J}$ cells, $G$ is the gravitational constant, and the isothermal sound speed $c_\text{s}$ in the dense molecule-forming regions is given by Equation~\ref{eq:snd_spd}. Thus, the relation between the number density $n$ and the spatial resolution $\Delta x$ for a simulation is
		\begin{equation}
			\label{eq:n_Jeans}
			n = \frac{\rho_\text{Jeans}}{\mu \, m_\text{p}} = \frac{\pi k_\text{B} T}{(\mu \, m_\text{p} \, N_\text{J} \, \Delta x)^2 G} \, .
		\end{equation}
	This leads to an expression for the required spatial resolution as a function of the gas number density, gas composition, and  temperature
		\begin{flalign}
			\label{eq:dx_numdens_relation}
			\Delta x &= \frac{1}{\mu \, m_\text{p} \, N_\text{J}}\sqrt{\frac{\pi k_\text{B} T}{n\,G}} \, , \nonumber \\
			&= 50\text{ pc} \frac{1}{\mu \, N_\text{J}} \left( \frac{T}{10\text{ K}} \right)^{1/2} \left( \frac{n}{1\text{~\pcmc}} \right)^{-1/2} \, .
		\end{flalign}
	For simplicity, $N_\text{J}=1$ is used for the conversion between the number density and spatial resolution.

\subsubsection{Resolution requirement for H$_2$ convergence}
 Substituting the density from Equation~\ref{eq:nH2_physical} in Equation~\ref{eq:dx_numdens_relation}, and taking $\mu=1.27$, $ N_\text{J}=1$, and $T=100$~K for the H$_2$ forming cold atomic gas, the required spatial resolution to satisfy the physical condition for H$_2$ formation is
	\begin{equation}
			\Delta x_{\text{H}_2,\text{physical}} \lesssim 40\text{ pc}\, .
	\end{equation}
Therefore, a low spatial resolution of $\sim40$~pc is enough to obtain densities for which H$_2$ formation from the warm neutral medium is faster than H$_2$ dissociation.

Similarly, for the dynamical condition, substituting Equation~\ref{eq:nH2_dynamical} in Equation~\ref{eq:dx_numdens_relation}, the expression for H$_2$ is
		\begin{equation}
			\label{eq:resolution_req_H2_exp}
			\Delta x_{\text{H}_2,\text{dynamical}} \leq 2\text{ pc} \left( \frac{1}{\mu \, N_\text{J}} \right)^2 \left( \frac{T}{10\text{ K}} \right)   \left(\frac{\langle \sigma \rangle_\text{mass}}{\text{1 km s}^{-1}}\right)^{-1}\, .
		\end{equation}	
	The molecule formation is effective for the dense gas with $n>10$~\pcmc (see Section~\ref{sec:H2_form_dissoc_time}), that has a mean velocity dispersion of  $\langle \sigma \rangle_\text{mass} \sim 2$~\kms in the CF runs (see Appendix~\ref{sec:appendix_veldisp}). Moreover, in the dense regions where the dynamical condition is important, $T=10$~K and $\mu \in [1.27, 2.35]$ depending on whether the H$_2$ forming region is locally predominantly atomic or molecular. This results in the required spatial resolution of
	\begin{equation}
		\label{eq:resolution_req_H2}
		\Delta x_{\text{H}_2,\text{dynamical}} \lesssim
		\begin{cases}
			& 0.6 \text{~pc~~~~for~} \mu=1.27 \\
			& 0.2 \text{~pc~~~~for~} \mu=2.35
		\end{cases}
	\end{equation}

\subsubsection{Resolution requirement for CO convergence}
Substituting the density from Equation~\ref{eq:nCO_physical} in Equation~\ref{eq:dx_numdens_relation}, and taking $\mu=2.35, \, N_\text{J}=1, \,\text{and~}T=10\text{~K}$ for the CO forming molecular (H$_2$) gas, the required spatial resolution to satisfy the physical condition for CO formation is
	\begin{equation}
			\Delta x_{\text{CO},\text{physical}} \lesssim 0.6\text{~pc} .
	\end{equation}
Therefore, a fairly high spatial resolution is required to obtain densities for which CO formation is faster than CO dissociation. 

	Similarly, for the dynamical condition, Equations~\ref{eq:nCO_dynamical}~and~\ref{eq:dx_numdens_relation} result in
		\begin{equation}
			\label{eq:resolution_req_CO_exp}
			\Delta x_{\text{CO},\text{dynamical}} \leq 0.5\text{ pc} \left( \frac{1}{\mu \, N_\text{J}} \right)^2 \left( \frac{T}{10\text{ K}} \right)   \left(\frac{\langle \sigma \rangle_\text{mass}}{\text{1 km s}^{-1}}\right)^{-1}\, .
		\end{equation}
	With $\langle \sigma \rangle_\text{mass} \sim 2$~\kms and $\mu$, $N_\text{J}, T$ as before, the required spatial resolution for converged CO formation is:
	\begin{equation}
		\label{eq:resolution_req_CO}
		\Delta x_{\text{CO},\text{dynamical}} \lesssim 0.04\text{~pc}\, .		
	\end{equation}
	
	For both H$_2$ and CO, the resolution requirement from the dynamical condition is more strict. These two derived resolution requirements are in agreement with the simulation results of H$_2$ and CO convergence in Section~\ref{sec:results_collflow}. For the typical velocity dispersion $\langle \sigma \rangle_\text{mass} \sim 2$~\kms, the cell crossing time $t_\text{cell,cross}$ for two different resolution runs is shown by the dot-dashed lines in Figures~\ref{fig:2dpdf_time_dens_H2}~and~\ref{fig:2dpdf_time_dens_CO} (cyan: CF-R6 and green: CF-R8). The coloured, vertical lines denote the densities corresponding to the spatial resolution of the two runs (Equation~\ref{eq:n_Jeans} with $\mu=2.35$, $N_\text{J}=1$,  and $T=10$~K).  Figure~\ref{fig:2dpdf_time_dens_H2} shows that the CF-R6 run with $\Delta x=0.125$~pc is able to fulfill both the physical and dynamical condition for H$_2$ formation. Similarly, Figure~\ref{fig:2dpdf_time_dens_CO} shows that the CF-R8 run with $\Delta x=0.032$~pc is able to fulfill both conditions for CO formation. Note that the density corresponding to the CF-R8 run (vertical, green line) is also comparable to the density at which the PDF begins to deviate from that of the CF-R9 and CF-R10 runs in Figure~\ref{fig:CF_1dpdf_dens}.

	\begin{figure}
		\includegraphics[width=\columnwidth]{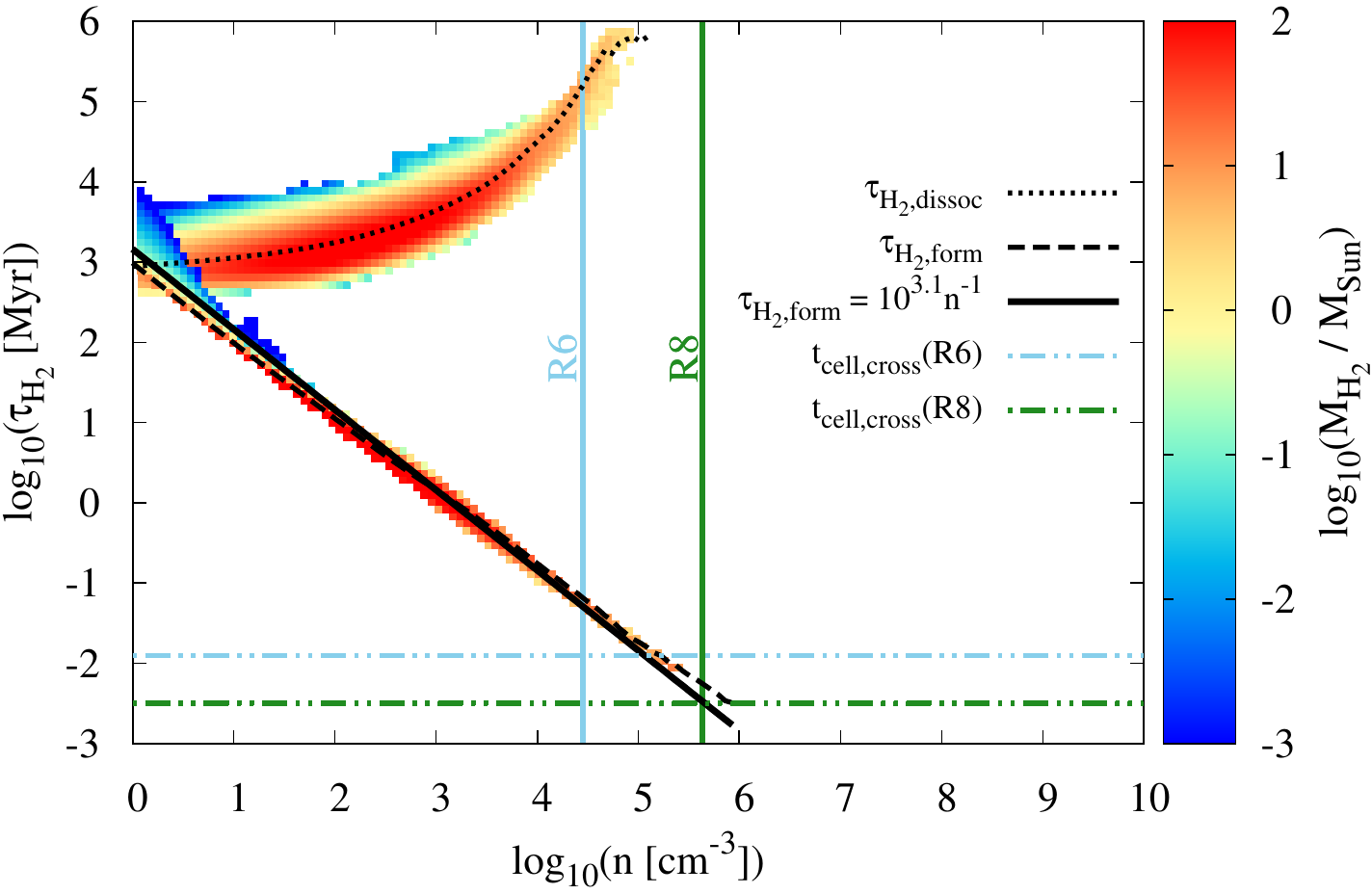}
		\caption{As in Figure~\ref{fig:2dpdf_time_dens_H2}, but now for the TB-R7-n030 run at $t=20$~Myr. The cell crossing time denoted by the horizontal dot-dashed lines are quite low since the velocity dispersion in the TB runs is high.}
		\label{fig:TB_2dpdf_time_dens_H2}
	\end{figure}

\subsection{Resolution requirements for the TB simulations} \label{sec:TB_res_req}
	In contrast to the CF runs, the TB runs have uniform resolution and self-gravity is activated only after 10 Myr. Thus the conversion from the densities to the length-scale as for the CF runs is not straightforward. Here, we focus on the resolution requirement for converged H$_2$ formation.
	
	 Figure~\ref{fig:TB_2dpdf_time_dens_H2} shows that the H$_2$ formation time-scale ($\tau_{\text{H}_2,\text{form}}$) in the TB-n030-R7 run at 20~Myr is similar to that seen in the CF runs (Figure~\ref{fig:2dpdf_time_dens_H2}). Therefore, the density expression in Equation~\ref{eq:nH2_dynamical} is valid for the TB runs as well. However, there is a notable difference in the density at which the physical condition (Equation~\ref{eq:physical_condt}) is satisfied because of the significantly different distribution of the dissociation time-scale ($\tau_{\text{H}_2,\text{dissoc}}$). The TB-R7-n030 run starts with cold atomic gas ($n_0=30$~\pcmc) and Figure~\ref{fig:TB_2dpdf_time_dens_H2} shows that the physical condition is satisfied for $n>1$~\pcmc. The coloured, horizontal, dot-dashed lines denote $t_\text{cell,cross}$ (Equation~\ref{eq:t_cell_cross}, with $\langle \sigma \rangle_\text{mass} = v_{\rm rms} =10$~\kms by construction) corresponding to two different resolutions -- cyan: R6 and green: R8. The coloured, vertical, solid lines refer to the density corresponding to the spatial resolution of R6 and R8, as in Figure~\ref{fig:2dpdf_time_dens_H2}. It is evident that the TB-R6-n030 run is no longer sufficient to fulfill the dynamical condition. Instead, the distribution indicates that refinement level 8 or higher would be required to fulfill both conditions for the H$_2$ molecule.
	 
	  For the gravitating gas in the TB-R*-n030 runs at $t=20$~Myr, Equation~\ref{eq:resolution_req_H2_exp} can be employed to obtain the necessary spatial resolution for converged H$_2$ convergence. For $\langle \sigma \rangle_\text{mass}=10$~\kms, this results in
	\begin{equation} \label{eq:resolution_req_H2_TB}
		\Delta x_{\text{H}_2,\text{dynamical}} \lesssim
		\begin{cases}
			& 0.1 \text{~pc~~~~for~} \mu=1.27 \\
			& 0.04 \text{~pc~~~~for~} \mu=2.35
		\end{cases}
	\end{equation}
	Such high resolution TB runs are computationally very demanding. Equation~\ref{eq:resolution_req_H2_exp} indicates that the convergence of H$_2$ chemistry is possible at lower resolutions if the velocity dispersion is lowered (i.e. if $t_\text{cell,cross}$ is increased in Figure~\ref{fig:TB_2dpdf_time_dens_H2}). This can be easily tested in the TB runs since the required $\langle \sigma \rangle_\text{mass}$ is a tuneable parameter of the setup. 
	
	The study of  multiple velocity dispersion values is quite expensive if the usual (32~pc)$^3$ box is used. Therefore, a smaller (8~pc)$^3$ simulation box is taken to carry out the TB-n030 runs with various $\langle \sigma \rangle_\text{mass}$ values (see Table~\ref{tab:runs}) and with resolutions corresponding to the TB-n030-R[5,6,7] runs. These test runs evolve for $t=10$~Myr without gravity. There is effectively less mass and less shielding capacity of the gas, and this affects the physical condition (Equation~\ref{eq:physical_condt}). However, the physical condition is still less restrictive than the dynamical condition. Thus, these runs are still valid to test the resolution requirement with various strengths of the supersonic turbulence in the gas. 
	
\begin{figure}
		\includegraphics[width=\linewidth]{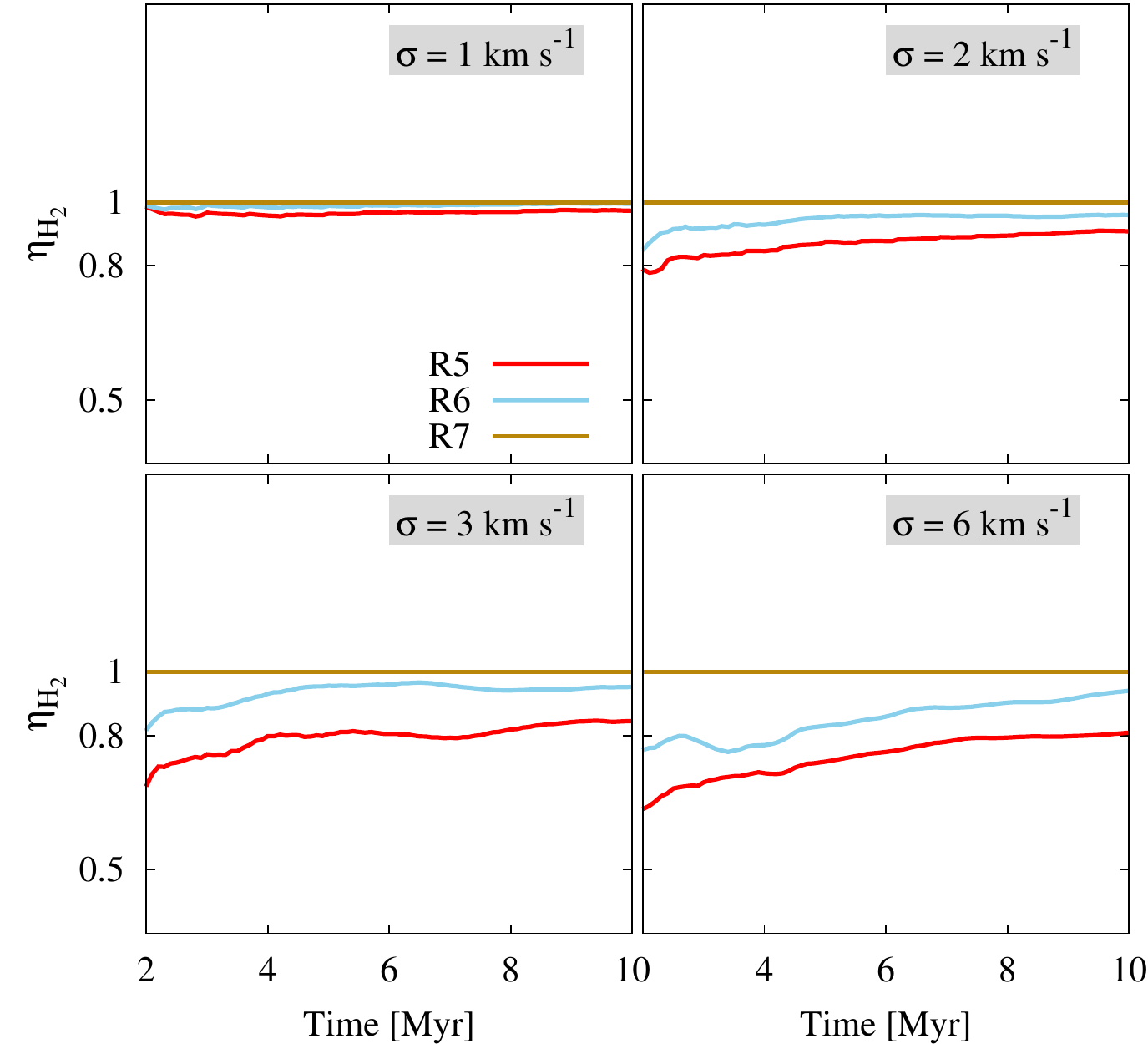}
		\caption{Evolution of $\eta_{\text{H}_2}$ (Equation~\ref{eq:etaH2}) for the TB runs with various strengths of the supersonic turbulence (runs TB-R[5,6,7]-n030-v[1,2,3,6], see Table~\ref{tab:runs}). As the 3D velocity dispersion increases, $\eta_{\text{H}_2}$ deviates a lot from $\eta_{\text{H}_2}=1$.}
		\label{fig:TB_evol_H2_veldisp}
	\end{figure}

	 Figure~\ref{fig:TB_evol_H2_veldisp} shows the evolution of $\eta_{\text{H}_2}$ (Equation~\ref{eq:etaH2}) in the test runs. It is evident that the low resolution simulations, that have converged evolution of H$_2$ content at low  $\langle \sigma \rangle_\text{mass}$, are not able to show convergence as the 3D velocity dispersion increases. The results can be summarized as follows:
	\begin{itemize}
		\item H$_2$ convergence is possible even for the TB-n030-R5 run with $\Delta x=0.250$~pc if $\langle \sigma \rangle_\text{mass} = 1$~\kms.
		\item The TB-n030-R6 run with $\Delta x=0.125$~pc is almost converged, with $\eta\gtrsim80$\% for $\langle \sigma \rangle_\text{mass}=2$~\kms or 3~\kms.
		\item When $\langle \sigma \rangle_\text{mass}=6$~\kms, $\eta_{\text{H}_2}$ is significantly lower than 1 for the TB-n030-R[5,6] runs.
	\end{itemize}
	Equation~\ref{eq:resolution_req_H2_exp} requires $\Delta x_{\text{H}_2,\text{dynamical}} \lesssim $ 0.4, 0.2, 0.12, and 0.06~pc for $\langle \sigma \rangle_\text{mass} = $ 1, 2, 3, and 6~\kms, respectively. Therefore, the numerical result from the test runs is consistent with the spatial resolution given by Equation~\ref{eq:resolution_req_H2_exp} for various $\langle \sigma \rangle_\text{mass}$ in the TB runs.
	
	In addition, Figure~\ref{fig:TB_1dpdf_dens} shows that the density PDFs in the pre-gravity phase are approximately log-normal, and higher resolution runs sample the high density regions more. According to Equation~\ref{eq:nH2_dynamical}, simulations with higher $\langle \sigma \rangle_\text{mass}$ require higher densities to satisfy the dynamical condition. Therefore, if the spatial resolution is high enough to sufficiently sample the critical densities given by Equation~\ref{eq:nH2_dynamical}~and~\ref{eq:nCO_dynamical}, molecule formation converges. This holds even if Equation~\ref{eq:dx_numdens_relation}, which converts a density to a spatial resolution using the Jeans condition, is not applicable.
	
\section{Conclusion}
	\label{sec:conclusions}
Recently, the inclusion of chemical networks has become more and more common in hydrodynamical simulations of the interstellar medium. However, a comprehensive study on the spatial resolution required to model different molecules is missing.
We present 3D hydrodynamic simulations coupled with a non-equilibrium chemistry network carried out using the MHD code \textsc{Flash4}. The effects of gas self-gravity, turbulence, and diffuse radiative transfer are taken into account. The chemical evolution under solar neighbourhood conditions is studied in two prevalent molecular cloud formation scenarios, (a) a turbulent periodic box and (b) a colliding flow. A large set of simulations with spatial resolution ranging from $\Delta x = 0.5$~pc to $\Delta x = 0.063$~pc for the turbulent box and up to $\Delta x = 0.008$~pc for the colliding flow simulations are performed to investigate the convergence of the evolving total H$_2$ and CO content. For the turbulent boxes we also investigate different initial mean number densities of 3, 30, and 300 cm$^{-3}$.

In all simulations, the morphology of the gas changes significantly with increasing resolution on both small and large scales. Typically, molecules form initially more slowly in low resolution runs. After some time, however, large clumps of gas form at low resolutions whereas numerous smaller, filamentary structures form at high resolutions. As the (self-)shielding in the large clumps becomes more effective, the low resolution runs eventually produce more H$_2$ and CO than the higher resolution runs. The simulations suggest that the resolution required to obtain the convergence of \textit{simple} quantities such as the total H$_2$ and CO content is high, even in quiescent environments subject to moderate levels of turbulence. It follows that more complex simulations with e.g. magnetic fields, high radiation fields, or supernova feedback potentially demand an even higher spatial resolution for the convergence of various chemical species than the one derived here.

The resolution requirements we derive follow from the understanding that the simulation needs to resolve typical densities at which (1) the molecule formation and the dissociation time scale are equal, which we refer to as a physical condition; and (2) the molecule formation and the cell crossing time scale are equal, which we refer to as a dynamical condition. The dynamical condition is more restrictive than the physical condition for both molecules. It requires the simulation to resolve 
	\begin{equation}
		n_{\text{H}_2,\text{dynamical}} \geq 10^{3.1}\text{ cm}^{-3} \left(\frac{\langle \sigma \rangle_\text{mass}}{\text{1 km s}^{-1}}\right)  \left( \frac{\Delta x}{\text{1 pc}}\right)^{-1} \, . \nonumber
	\end{equation}
		 for converged H$_2$ formation and
	\begin{equation}
		n_{\text{CO},\text{dynamical}} \geq 10^{3.7}\text{ cm}^{-3} \left(\frac{\langle \sigma \rangle_\text{mass}}{\text{1 km s}^{-1}}\right)  \left( \frac{\Delta x}{\text{1 pc}}\right)^{-1} \, . \nonumber
	\end{equation}
		 for converged CO formation. We find that the required spatial resolution depends on the composition, temperature, and the strength of turbulence of the MC forming dense gas. In particular, we derive that a spatial resolution required  in the colliding flow setup is
\begin{enumerate}
	\item $\Delta x \lesssim 0.2\text{~pc}$ to model H$_2$ formation
	\item $\Delta x \lesssim 0.04\text{~pc}$ to model CO formation
\end{enumerate}

This is consistent with the numerical results of the colliding flow simulations, which are converged in H$_2$ and CO for these spatial resolutions.
	
In the driven turbulent box with a driving velocity of 10 km s$^{-1}$, neither the total H$_2$ content, nor the total CO content are converged when a spatial resolution up to $\Delta x=0.063$~pc is used. The driven turbulence in warm atomic gas severely affects the evolutionary history of the H$_2$ and CO molecules because any dense region capable of sustaining the molecular gas is quickly disrupted. Similar to the colliding flow runs, the evolution of the CO content is more severely affected by the resolution. We can explain why the molecule formation is unresolved in these runs: condition 2 is not fulfilled because the required spatial resolution scales with one over the mass-weighted velocity dispersion, and therefore a factor of 5 higher resolution would be required to model the formation of H$_2$ and CO. 

\section*{Acknowledgements}
The authors are grateful to the anonymous referee for the insightful comments that strengthened the paper. The authors are grateful to J. C. Ib{\'a}{\~n}ez-Mej{\'i}a and F. Dinnbier for fruitful discussions. PRJ, SW, and DS acknowledge the support from the DFG through SFB 956 sub-project C5. SW and SDC acknowledge funding by the European Research Council through the ERC Starting Grant ``RADFEEDBACK: The radiative ISM" (project number 679852). SW and SCOG acknowledge the support of the DFG via the SPP 1573, ``Physics of the Interstellar Medium'' (grant number GL 668/2-1). SCOG acknowledges support from the DFG via SFB 881, ``The Milky Way System'' (sub-projects B1, B2 and B8). SCOG also acknowledges support from the European Research Council under the European Community's Seventh Framework Programme (FP7/2007 - 2013) via the ERC Advanced Grant ``STARLIGHT: Formation of the First Stars'' (project number 339177). The software used in this work was in part developed by the DOE NNSA-ASC OASCR Flash Center at the University of Chicago\footnote{FLASH is available at \url{http://www.flash.uchicago.edu/}}. The authors gratefully acknowledge the Gauss Centre for Supercomputing e.V. 
(\url{www.gauss-centre.eu}) for funding this project by providing computing time on the GCS Supercomputer
SuperMUC at Leibniz Supercomputing Centre (\url{www.lrz.de}) and thank the Regional Computing Center of the University of Cologne (RRZK) for providing computing time on the DFG-funded High Performance Computing (HPC) system CHEOPS as well as support. The authors are also thankful to \textsc{VisIt} \citep{HPV:VisIt} which allowed to produce the volume-rendered images.



\bibliographystyle{mnras}
\bibliography{astro}



\appendix
\section{The velocity dispersion of the dense gas in Colliding Flow setup}
	\label{sec:appendix_veldisp}
\begin{figure}
		\includegraphics[width=\columnwidth]{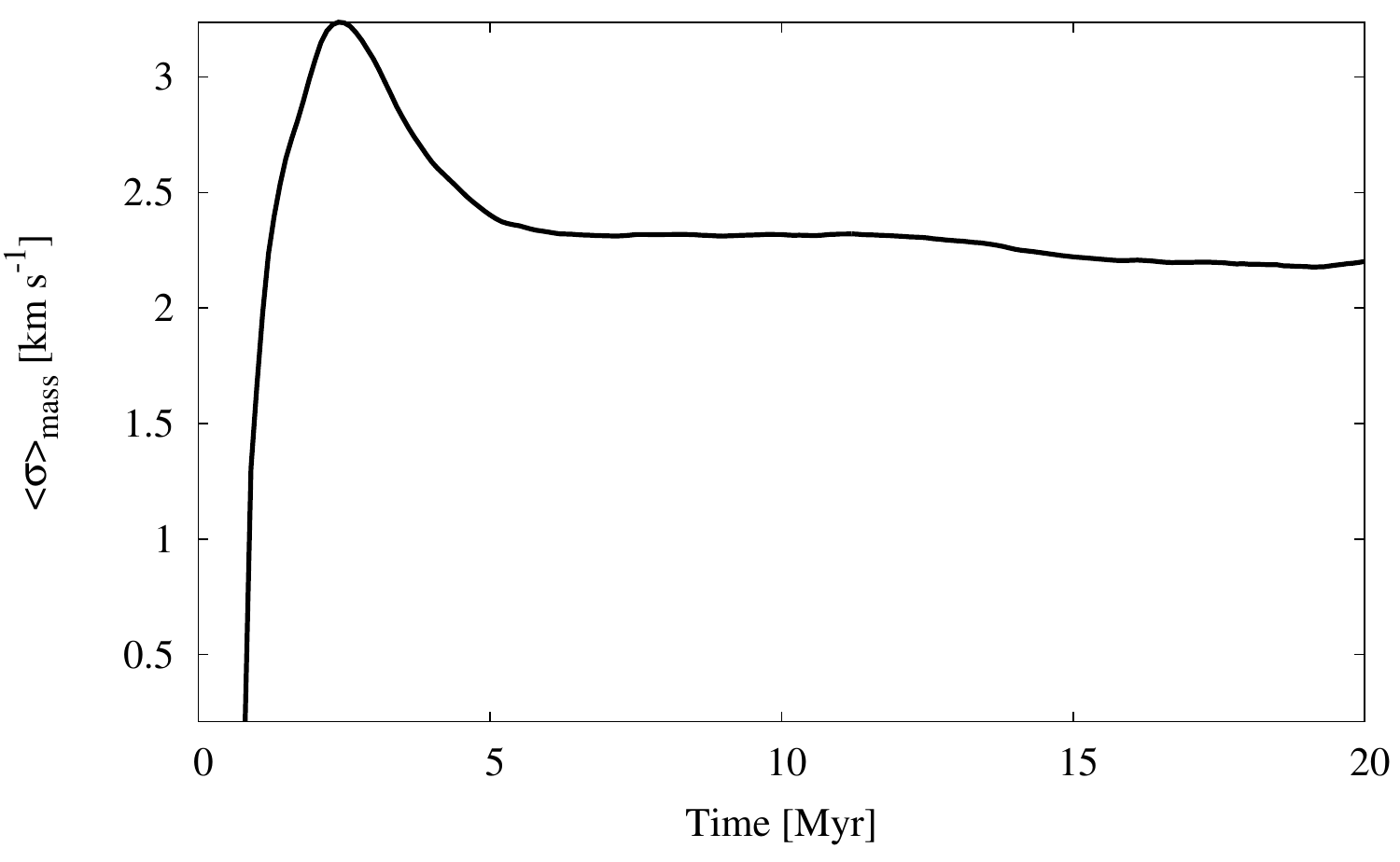}
		\caption{The evolution of the mass-weighted mean velocity dispersion $\langle\sigma\rangle_\text{mass}$ in the CF-R10 run. Only the cells with density $n>10$~\pcmc, that fulfill the physical condition, is considered. Such gas has typical $\langle\sigma\rangle_\text{mass}\sim 2$~\kms maintained throughout the molecular gas formation period.}
		\label{fig:CF_veldisp}
	\end{figure}

	Figure~\ref{fig:CF_veldisp} shows the evolution of the mass-weighted mean velocity dispersion of the dense gas with number density $n>10$~\pcmc for the CF-R10 run. This dense gas contains $\sim90\%$ of the total mass of the gas with $T<4000$~K. After $t=5$~Myr, the velocity dispersion of $\sim2$~\kms is maintained during the cloud evolution as the dense regions become progressively more massive.
	
\section{The molecular content in Jeans unresolved regions}
	\label{sec:appendix_Jeans_unresolved}
\begin{figure}
		\includegraphics[width=\columnwidth]{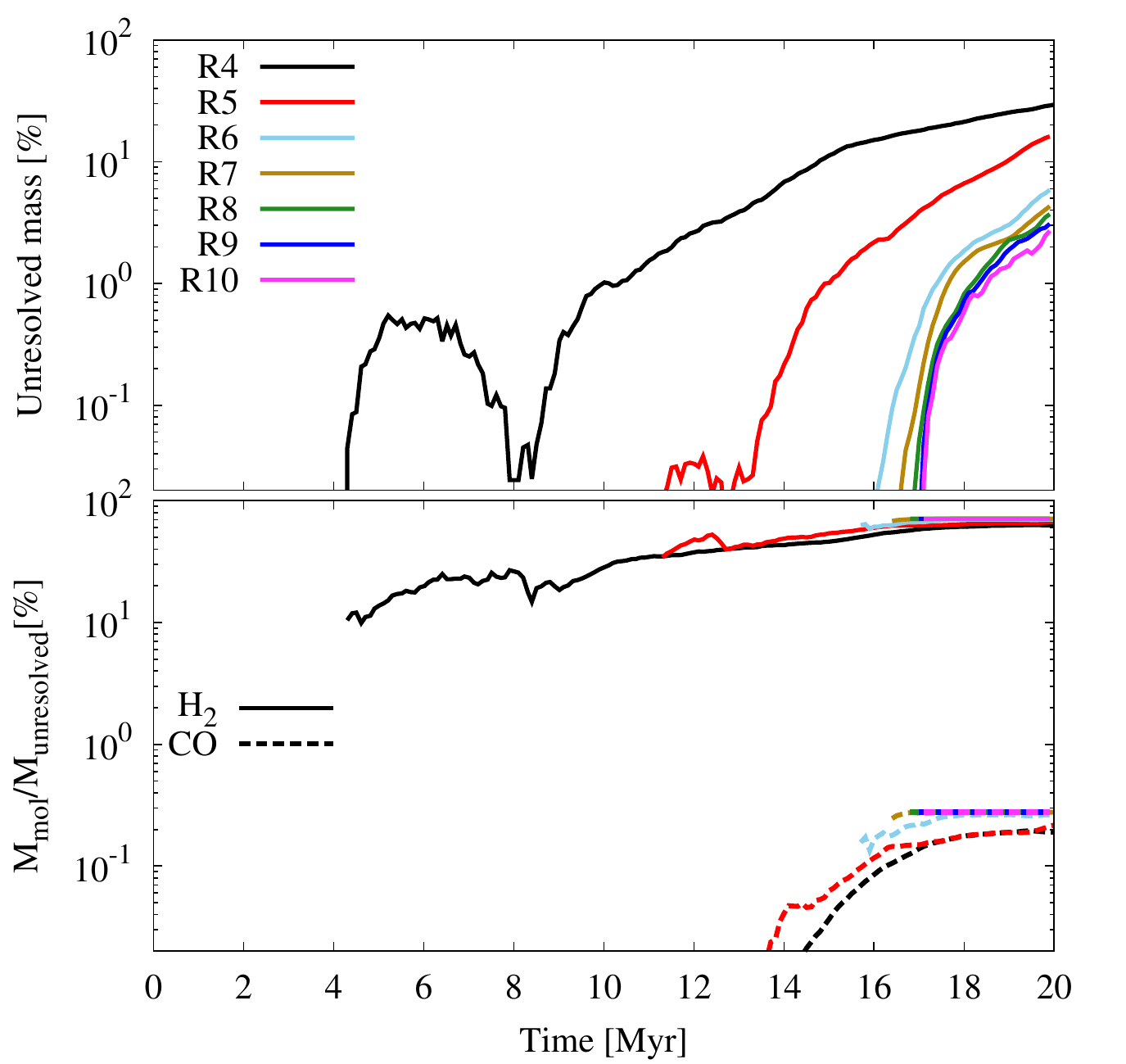}
		\caption{The evolution of (top) the percentage of the mass in the Jeans unresolved regions of the CF runs and (bottom) the percentage of the mass of H$_2$ (solid lines) and CO (dashed lines) in such unresolved regions. In the high resolution runs (CF-R6 and higher), the unresolved regions contain $\sim3-6$\% of the total gas mass, that in turn have $\sim70$\% of their mass in H$_2$ and CO.}
		\label{fig:CF_evol_unresolved}
	\end{figure}

	Figure~\ref{fig:CF_evol_unresolved} shows the evolution of (top) the percentage of the mass contained in the Jeans unresolved regions of the CF runs and (bottom) the amount of molecular gas (H$_2$ and CO) in such unresolved regions. With increasing resolution, the mass of Jeans unresolved regions decrease significantly, i.e. $\sim30$\% in CF-R4 to $\sim5$\% in CF-R6 and higher resolution runs. For all resolutions, such dense regions hold $\sim70$\% of their mass in H$_2$ and CO molecules. Therefore, most of the molecular gas investigated for the resolution criteria are contained in resolved density structures.
	
\bsp	
\label{lastpage}
\end{document}